\documentclass[12pt]{article}
\usepackage{scicite}
\usepackage{graphicx}
\usepackage{times}
\usepackage{amssymb}

\newenvironment{sciabstract}{%
\begin{quote} \bf}
{\end{quote}}

\newcounter{lastnote}
\newenvironment{scilastnote}{%
\setcounter{lastnote}{\value{enumiv}}%
\addtocounter{lastnote}{+1}%
\begin{list}%
{\arabic{lastnote}.}
{\setlength{\leftmargin}{.22in}}
{\setlength{\labelsep}{.5em}}}
{\end{list}}

\title{Black hole lightning due to particle acceleration at subhorizon scales\\
\vspace*{0.5cm}
\normalsize{\textit{Science}, submitted May 16; accepted October 23, 2014}
}

\author
{ 
J. Aleksi\'c$^{1}$, S. Ansoldi$^{2}$, L. A. Antonelli$^{3}$, P. Antoranz$^{4}$, A. Babic$^{5}$, P. Bangale$^{6}$,\\
 J. A. Barrio$^{7}$, J. Becerra Gonz\'alez$^{8,25}$,  W. Bednarek$^{9}$, E. Bernardini$^{10}$
% ...
\\
\normalsize{$^{1}$ for affiliations see list below}\\
\\
\normalsize{$^\ast$To whom correspondence should be addressed;}\\
\normalsize{E-mail: Dorit Eisenacher: deisenacher@astro.uni-wuerzburg.de;}\\
\normalsize{Julian Sitarek: jsitarek@ifae.es, Karl Mannheim: mannheim@astro.uni-wuerzburg.de}\\
}

\date{}

\begin{document} 

\baselineskip24pt

\maketitle 

\begin{sciabstract}

Supermassive black holes with masses of millions to billions of solar masses are commonly found in the centers of galaxies. 
Astronomers seek to image jet formation using radio interferometry, but still suffer from insufficient angular resolution. 
An alternative method to resolve small structures is to measure the time variability of their emission. 
Here, we report on gamma-ray observations of the radio galaxy IC~310 obtained with the MAGIC telescopes revealing variability with doubling time scales faster than 4.8 min. 
Causality constrains the size of the emission region to be smaller than 20\% of the gravitational radius of its central black hole. 
We suggest that the emission is associated with pulsar-like particle acceleration by the electric field across a magnetospheric gap
at the base of the radio jet. 
\end{sciabstract}

More than three decades ago it was proposed that the radio emission of extragalactic jets 
results from a relativistically moving plasma consisting of magnetic fields and accelerated particles following a power-law energy distribution \cite{blandford79}.
One of the major assets of the model is that it can explain the non-thermal emission of extragalactic jets across the entire electromagnetic
spectrum, from radio waves up to gamma rays.  The emission can be understood as
synchrotron radiation and inverse Compton scattering \cite{maraschi92, dermer}
due to particles
accelerated at shock waves in the jets.
The gamma rays can reach very high energies measured in Giga-electronvolts ($1$~GeV$=10^9$~eV corresponding roughly to the rest mass energy 
equivalent of the proton) and Tera-electronvolts ($1$~TeV$=10^{12}$~eV).
According to the Blandford-Znajek mechanism, the jets are powered by extracting rotational energy
from the black holes which have acquired angular momentum through the accretion of surrounding gas and black hole mergers \cite{blandfordznj77},
although so far astrophysical evidence for the role of black hole spin is still lacking \cite{vanvelzen13}.
For a maximally rotating supermassive black hole of mass $M=10^8 m_8 M_\odot$, where $M_\odot$ denotes one solar mass, 
the size of the jet formation region should be of the order of its
gravitational radius $r_{\rm g}=G_{\rm N}M/c^2\sim 1.5\times 10^{11}\, m_8$~m and twice this value for a non-rotating Schwarzschild black hole.
Astronomical telescopes do not yet provide the angular resolution needed to image structures on this scale.
The highest resolution images of jets obtained with very long baseline radio interferometry show
radio-emitting knots traveling down the jets \cite{marscher}. 
Approaching the black hole, the spectra cut off at increasingly higher frequencies due to synchrotron-self-absorption.
Observations at very high frequencies where the core becomes transparent are
needed to zoom into the region where the jets are emerging from. 
The record holder is a very long baseline radio interferometry observation of the jet of the nearby radio galaxy M87
at a frequency of 230 GHz, resolving a source with the size of $11.0\pm0.8$ gravitational radii \cite{doeleman12}.

{\bf The event horizon light-crossing time.} 

Whilst direct imaging of the jet formation region has to await better angular resolution, 
indirect information about its size can be inferred from the temporal variability of the emission coming from that region.
The observed gamma-ray variability time scales indeed reach down to the 
event horizon light crossing time $\Delta t_{\rm BH}=r_{\rm g}/c=G_{\rm N}M/c^3=8.3\, m_8$~minutes \cite{vovk13}
vindicating the scenario that the jets originate from the magnetospheres of accreting black holes. 
An example is the radio galaxy M87 in the Virgo cluster of galaxies\cite{albert08, abr12}.
This galaxy harbors a central supermassive black hole with the enormous mass of $\sim6.4\times 10^9M_\odot$ \cite{gebhardt09}.
M87 exhibits gamma-ray variability at a time scale of days \cite{acc09} which is consistent with the
light-crossing time of the event horizon $\Delta t_{\rm BH}(\rm M87)=0.4$~days.

{\bf Smaller than a black hole?}
  
Observations of Mrk\,501 \cite{albert07a} and PKS~2155$-$304 \cite{aharonian07} at very high energies have provided evidence for extreme variability events with 
flux doubling time scales as short as $\sim 2$~min. 
The ultrafast variability corresponds to $\Delta t< \Delta t_{\rm BH}$ and therefore casts a shadow of doubt on the current shock-in-jet paradigm.
It has been suggested that relativistic bulk motion of the jets could explain the observed time scales \cite{begelman08}.
The argument relies on the observation that these flaring sources belong to the class of sources that astronomers call blazars.
In blazars,  the jets are pointing at a small angle towards the observer.  Since the jet plasma moves with 
a speed close to the speed of light $\beta=v_{\rm j}/c \simeq 1$ leading to a bulk Lorentz factor $\Gamma_{\rm j}>1$, several effects arise due to the relativistic boosting 
of the emission \cite{notedoppler}.  One of them affects the time scale 
of flux variations of the emission from a shock. The moving shock plasma almost catches up with its own
radiation, and this leads to a shortening of the observed variability time scale $\Delta t$ compared with the
variability time scale $\Delta t'$ in a frame comoving with the shock given by 
$\Delta t=(1+z)\delta^{-1}\Delta t'$ where $z$ denotes the cosmological redshift of the source,
For a given value of $\Gamma_{\rm j}$, the Doppler factor $\delta$ depends strongly on the orientation angle of the
jet $\theta$  (Fig.~1).  Note that $\theta=0$ corresponds to perfect alignment.   
For Mrk\,501 and PKS\,2155$-$304, almost perfectly aligned jets with $\Gamma_{\rm j}>50$  
would be needed to accomodate for $\Delta t< \Delta t_{\rm BH}$ and to avoid self-absorption of the gamma rays due to pair production \cite{begelman08}.
In blazars, interferometric observations of the superluminal motion of radio knots 
suggest lower values $\Gamma_{\rm j}\sim 10$ and orientation angles of a few degrees \cite{lister09}. 
Larger values of $\Gamma_{\rm j}$ would  lead to a problem with population statistics:  the number of
unbeambed counterparts of blazars viewed at larger angles would then exceed the number of radio galaxies, 
commonly believed to represent the misaligned blazars \cite{urry91}.
Assuming lower black hole masses would bring down $\Delta t_{\rm BH}$.
However, lower masses conflict with the firmly established dynamical measurements of black hole masses, 
and therefore do not seem to be a likely solution of the dilemma.
Other possible solutions of this {\em Doppler factor crisis} \cite{lyutikov10} invoke models of structured jets \cite{ghisellini08, giannios09} or
Poynting flux dominated jets in which only a few but very fast seed 
particles at the jet base  reach high Lorentz factors \cite{michel69, lyutikov09, kirk11}, before the Poynting flux is 
converted into the kinetic energy of the bulk flow by mass entrainment.  

All of these attempts to explain the sub-horizon scale variability with relativistic projection effects alone ignore a fundamental problem. 
If the perturbations giving rise to the blazar variability are injected at the jet base, the time scale of the flux variations
in the frame comoving with the jet is affected by time dilation with Lorentz factor $\Gamma_{\rm j}$.  
In blazars where $\delta\sim\Gamma_{\rm j}$, the Lorentz factor cancels out, and the observed
variability time scale is ultimately bounded below by $\Delta t_{\rm BH}$. 

{\bf IC~310: A gamma-ray lightning inferno.} 

IC~310 is a peculiar radio galaxy located in the outskirts of the Perseus cluster at a distance of 260 million light years from Earth. 
On November 12/13$^{\rm th}$ in 2012, MAGIC, a system of two Imaging Atmospheric Cherenkov Telescopes located on the Canary island of La Palma \cite{aleksic12},
detected an extraordinary outburst of gamma rays from this object (see Fig.~2, S1).
The details of the analysis can be found in Section~S2. 
Prior to these observations, variable gamma-ray emission from IC\,310 had already been detected by satellite and ground-based gamma-ray
instruments at GeV and TeV energies, e.g. \textit{Fermi}-LAT and MAGIC \cite{neronov10, aleksic10, aleksic13}.
In the night of the flare in November 2012, the mean flux above 300\,GeV was $(6.08 \pm 0.29)\times10^{-11}$\,cm$^{-2}$s$^{-1}$, i.e., 
four times higher than the highest flux
during previous observations in 2009/2010. 
The measured spectrum (Fig.~3) can be described by a simple power law with a differential photon spectral index of 
$\Gamma=1.90\pm0.04_{\mathrm{stat}}\pm0.15_{\mathrm{syst}}$ in the energy range of 70\,GeV - 8.3\,TeV (see Table S2).
Owing to its proximity, the spectrum of IC\,310 is only marginally affected by photon-photon absorption in collisions with the extragalactic background light (EBL).

IC~310 harbors a supermassive black hole with a mass of $M=(3^{+4}_{-2}) \times10^{8}M_\odot$ (see Section~S1.1) corresponding to an event horizon light-crossing 
time of $\Delta t_{\rm BH}=(23^{+34}_{-15})$~min.
The mass has been inferred from the correlation of black hole masses with the central velocity dispersion 
of their surrounding galaxies \cite{gultekin2009, mcelroy95}.  
The reported errors are dominated by the intrinsic scatter of the distribution.  
The same value of the mass is obtained from the fundamental plane of black hole activity \cite{merloni03}.
The scatter in the fundamental plane for a single measurement is larger and corresponds to a factor of $\sim 7.5$.

During 3.7\,hrs of observations, extreme variability with multiple individual flares has been detected (Fig.~4, S3, S4). 
The flare has shown the most rapid flux variations ever observed in extragalactic objects, comparable only to those seen in Mrk\,501 
and PKS\,2155$-$304.   A conservative estimate of the shortest variability time scale in the frame of IC\,310 yields $\Delta t/(1+z)=4.8$~min.
It is the largest doubling time scale with which the rapidly rising part of the flare can be fitted with a probability $>5\%$ (see Fig.~S4).
The light curve also shows pronounced large-amplitude flickering characterized by doubling time scales down to $\Delta t\sim 1$~min.
The conservative variability time scale corresponds to $20\%$ of the light travel time across the event horizon, or still $60\%$ of it giving allowance
for the scatter in the dynamical black hole mass measurement.

From the absence of a counter radio jet and the requirement that the proper jet length does not exceed the maximum of the 
distribution of jet lengths in radio galaxies, the orientation angle was found to be in the range 
$\theta\sim$10$^{\circ}$--20$^{\circ}$ (see Section~S1.2), and the Doppler factor consistent with 
$\delta\approx 4$ \cite{kadler12}. These values put IC~310 at the borderline between radio galaxies and blazars. 
The jet power estimated from observations of the large-scale radio jet is $L_{\rm j}=2\times 10^{42}$~erg~s$^{-1}$ 
assuming that it contains only electrons, positrons, and magnetic fields in equipartition of their energy densities (see Section~S1.3).  
For a radiative efficiency of $10\%$, the Doppler-boosted average luminosity of the jet emission amounts to $0.1\delta^4 L_{\rm j}\approx 5\times 10^{43}$~erg~s$^{-1}$
which is close to the one observed in very-high energy gamma rays.  
For $\delta\sim 4$, the variability time scale in the comoving frame of the jet, where it should be larger than $\Gamma_{\rm j}\Delta t_{\rm BH}$,  is actually close to $\Delta t_{\rm BH}$ (cf. Fig.1).   A very high value of the  Doppler factor is required to avoid the absorption of the gamma rays due to interactions with low-energy synchrotron photons, inevitably co-produced with the gamma rays in the shock-in-jet scenario. The optical depth to pair creation by the gamma rays can be approximated by
$\tau_{\gamma\gamma}(10~\rm TeV) \sim 300 \left(\delta / 4\right)^{-6}\left (\Delta t / 1~min\right)^{-1} \left(L_{\rm syn} / 10^{42}~\rm erg~s^{-1}\right)$.  
Adopting a non-thermal infrared luminosity of $\sim 1\%$ of the gamma-ray luminosity during the flare,
the emission region would be transparent to the emission of 10~TeV gamma rays only if $\delta\gtrsim10$.
For the range of orientation angles inferred from radio observations, the Doppler factor is constrained to a value of $\delta< 6$ (Fig.~1). 
One can speculate if the inner jet, corresponding to the unresolved radio core, bends into a just-right orientation angle 
to produce the needed high value of the Doppler factor (see Section~S1.2).  The probability for such an alignment seems to be rather low.
Moreover, the observed radio jet does not show any signs of a perturbation of its flow direction on the parsec and kiloparsec scales.
Since perturbations of the flow direction of the inner jet would later propagate to larger scales, major bends apparently never occurred in the past.

In summary, trying to interpret the data in the frame of the shock-in-jet model meets difficulties.   
Considering the role of time dilation
renders a solution of this problem impossible for any value of $\Gamma_{\rm j}$.
Therefore, we conclude that the observations indicate a sub-horizon scale emission region of a different nature.

{\bf Possible origins of sub-horizon scale variability.}
 Substructures smaller than the event horizon scale emitting highly anisotropic radiation 
(to avoid pair absorption) seem to be responsible for the minute-scale flux variations. 
The possible explanations fall into three categories:  (i) mini-jet structures within the jets \cite{giannios10}, 
(ii) jet-cloud interactions where the clouds may originate from stellar winds \cite{bednarek97, barkov10, barkov12}, and (iii) 
magnetospheric models \cite{rieger00, neronov07, neronov09, levinson11, beskin92}, similar to those known from pulsar theory.        

According to the mini-jet model (i), plasmoids resulting from magnetic reconnection traveling down the jet with a relativistic speed 
are responsible for the minute-scale flux variations observed in blazars. The model could help
to mitigate the constraints on the bulk Lorentz factor by introducing a larger effective bulk Lorentz factor for the plasmoids.
The mechanism also predicts reconnection events from regions outside of the beaming cone
$\sim \Gamma_{\rm j}^{-1}$ that could explain the day-scale flares from the radio galaxy M87 invoking external radiation fields as a target for inverse Compton scattering \cite{giannios10}.  
However,  the off-axis mini-jet luminosity depends on $(\Gamma_{\mathrm{j}}\theta)^{-8}$ and the jet power required for IC~310 is two orders of magnitude
higher compared to the one estimated from radio observations (see Section~S3.2). Thus, this model is challenged
by the observed high luminosity in IC~310 during the flares.

Substructures smaller than the jet radius may also be introduced by considering interactions between clouds and the jet (ii).  The original shock-in-jet model \cite{blandford79} 
considered this to be the main source of mass entrainment and predicted variability from the process.  Recently, more elaborate work on the model has had some success in
explaining the variability of M87 by pp-collisions due to the bombardment of clouds boiled off red giants with protons in the jet \cite{barkov12}.  However, the model is linked to the cloud crossing time of the jet and the 
proton-proton cooling time, both of which by far exceed the event horizon scale.  
Faster variability could be observed in case the cloud gets destroyed but a strong beaming effect would then be needed to explain the observed luminosities.  
In another variant, drift acceleration of particles along the trailing shock behind the stellar wind of a star interacting with the jet is considered.
This might lead to an extremely anisotropic emission pattern. 
As mass-loosing stars sweep across the jet, passing magnetic field lines pointing to the observer, 
the postulated accelerated particle beams in their trails become visible 
for a short time.  For IC~310, the emission 
would have to be confined to within an angle of $\alpha\sim10^{-5}$~rad to explain
the observed variability time scale, requiring a very stable direction of the accelerated particle
beams, at a large angle to the jet main thrust.  Since two-fluid particle beams are prone
to numerous plasma instabilities, the scenario relies on unphysical assumptions.

In magnetospheric models (iii), particle acceleration is assumed to be due to electric fields parallel to the magnetic fields. This mechanism is known to
operate in the particle-starved magnetospheres of pulsars, but it could also operate in the magnetospheres anchored to the ergospheres of accreting black holes (see Fig.~5).
Electric fields can exist in vacuum gaps  when the density of charge carriers is too low to warrant 
their shortcut.  The critical charge density for the vacuum gaps is the so-called Goldreich-Julian charge density.  
Electron-positron pairs in excess of the Goldreich-Julian charge density can be produced
thermally by photon-photon collisions in a hot accretion torus or corona surrounding the black hole.
It has also been suggested, that particles can be injected by the reconnection of twisted magnetic loops in the accretion flow
\cite{neronov09}.
A depletion of charges from thermal pair production is expected to happen when the accretion rate becomes very low.  In this late phase of their
accretion history,  supermassive black holes are expected to have spun up to maximal rotation.  
Black holes can sustain a Poynting flux jet by virtue of the Blandford-Znajek mechanism \cite{blandfordznj77}. 
Jet collimation takes place rather far away from the black hole at the scale of the light cylinder beyond $\sim 10r_{\rm g}$.
Gaps could be located at various angles with the jet axis corresponding to the polar and outer gaps in pulsar magnetospheres
leading to fan beams at rather large angles with the jet axis.
The gap emission must be highly variable, since gap height and seed particle content depend sensitively on plasma turbulence and
accretion rate. For an accretion rate of $\dot m\sim 10^{-4}$  (in units of the Eddington accretion rate) and maximal black hole rotation, 
the gap height in IC~310 is expected to be $h\sim 0.2 r_{\rm g}$ \cite{levinson11} which is in line with the observations.
Depending on the  electron temperature and geometry of the radiatively inefficient accretion flow, its thermal cyclotron luminosity 
can be low enough to warrant the absence of pair creation attenuation 
in the spectrum of gamma rays. In this picture, the intermittent variability witnessed in IC~310 is due to a runaway effect.  As particles accelerate to ultrahigh energies, 
electromagnetic cascades develop multiplying the number of charge carriers until their current shortcuts the gap.  
The excess particles are then swept away with the jet flow, until the gap reappears.

Radio galaxies and blazars with very low accretion rates allow us to obtain a glimpse of the jet formation process near supermassive black holes. 
The sub-horizon variability in combination with the results from direct imaging campaigns invite to explore analogies with pulsars where
particle acceleration takes place in two stages. In the first stage, particle acceleration occurs in the gaps of a charge-separated magnetosphere
anchored in the ergosphere of a rotating black hole, and in a second stage at shock waves in the force-free wind beyond the outer light cylinder.

\bibliographystyle{Science} 

\begin{scilastnote}
\item We would like to thank
the Instituto de Astrof\'{\i}sica de Canarias
for the excellent working conditions
at the Observatorio del Roque de los Muchachos in La Palma.
The support of the German BMBF and MPG,
the Italian INFN,
the Swiss National Fund SNF,
and the Spanish MICINN
is gratefully acknowledged.
This work was also supported
by the CPAN CSD2007-00042 and MultiDark CSD2009-00064 projects of the Spanish Consolider-Ingenio 2010 programme,
by grant 127740 of the Academy of Finland,
by the DFG Cluster of Excellence ``Origin and Structure of the Universe'',
by the Croatian Science Foundation (HrZZ) Projects 09/176,
by the University of Rijeka Project 13.12.1.3.02,
by the DFG Collaborative Research Centers SFB823/C4 and SFB876/C3,
and by the Polish MNiSzW grant 745/N-HESS-MAGIC/2010/0.
We thank also the support by DFG WI 1860/10-1.
J. S. was supported by ERDF and the Spanish MINECO through
FPA2012-39502 and JCI-2011-10019 grants.
E. R. was partially supported by the Spanish MINECO projects
AYA2009-13036-C02-02 and AYA2012-38491-C02-01 and by
the Generalitat Valenciana project PROMETEO/2009/104, as
well as by the COST MP0905 action 'Black Holes in a Violent
Universe'.
The European VLBI Network is a joint facility of European, Chinese,
South African and other radio astronomy institutes funded by their national research councils.
The research leading to these results has received funding from
the European Commission Seventh Framework Programme (FP/2007-2013)
under grant agreement No. 283393 (RadioNet3).
The MAGIC data is archived in the data center
at the Port d\'Informaci\'o Cient\'{\i}fica (PIC) in Barcelona.
The EVN data are available at the Data Archive at the
 Joint Institute for VLBI in Europe (JIVE).
\end{scilastnote}

\clearpage

 \begin{figure}
    \centering
       \includegraphics[width=10cm]{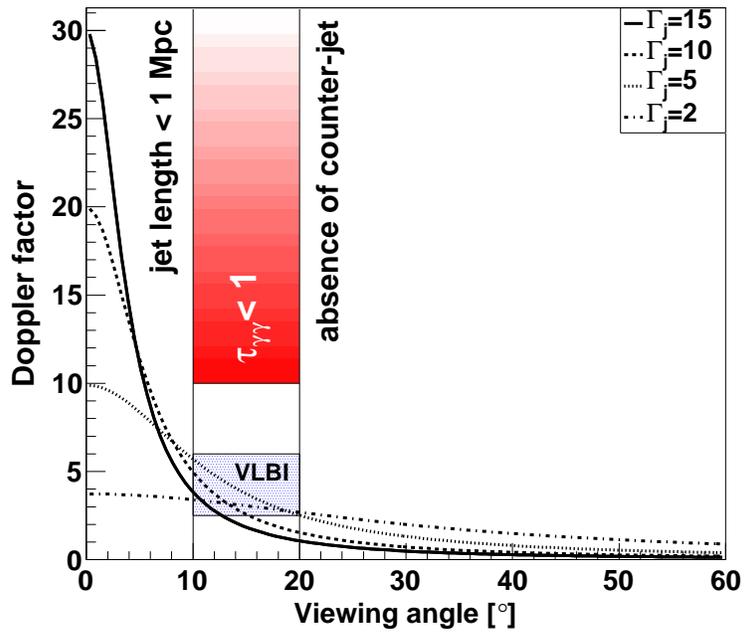}
      \caption{The relation between Doppler factor and orientation angle evaluated for various values of the bulk Lorentz factor compared with the observational constraints of these parameters.
 The blue box shows the constraints on the Doppler factor arising from radio observations \cite{kadler12}.  For illustrative purposes, the red box shows the constraint from the gamma-ray optical depth to pair creation assuming
 $L_{\rm syn}\sim 1\% L_{\rm VHE}$.
 }
               \label{Doppler}%
     \end{figure}

 \begin{figure}
    \centering
       \includegraphics[width=13cm]{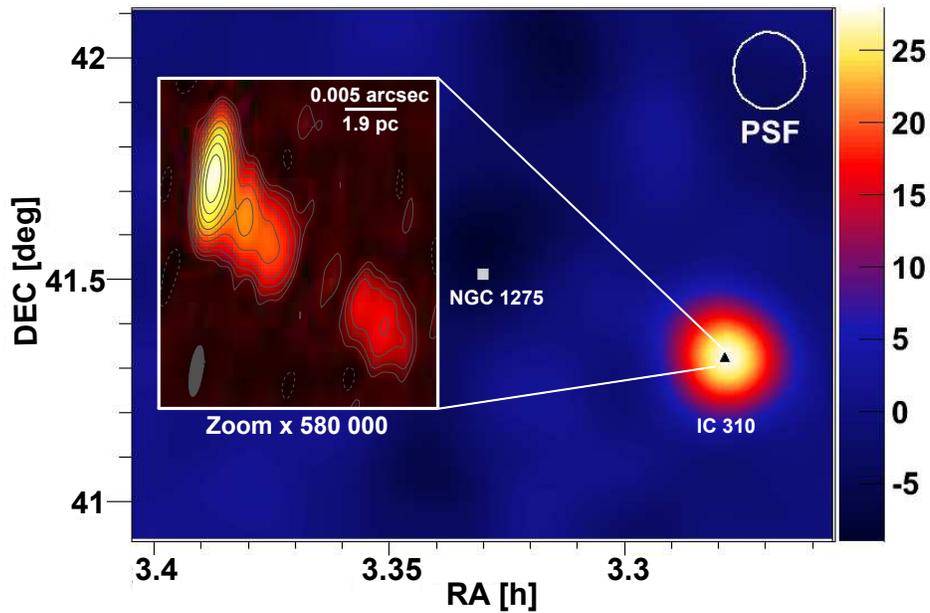}
      \caption{Significance map (color scale) of the Perseus cluster in gamma rays observed in the night of November 12/13$^{\mathrm{th}}$, 2012, with 
 the MAGIC telescopes. The inset shows the radio jet image of IC~310 at 5.0\,GHz obtained with the European VLBI Network (EVN) on
 October 29, 2012. Contour lines (and associated to them color scale) increase logarithmically by factors of
 2 starting at three times the noise level (see supplement for image parameters).
 The ratio of the angular resolution
 between MAGIC and the EVN is 1:580\,000.}
               \label{Skymap}%
     \end{figure}

 \begin{figure}
    \centering
       \includegraphics[width=15cm]{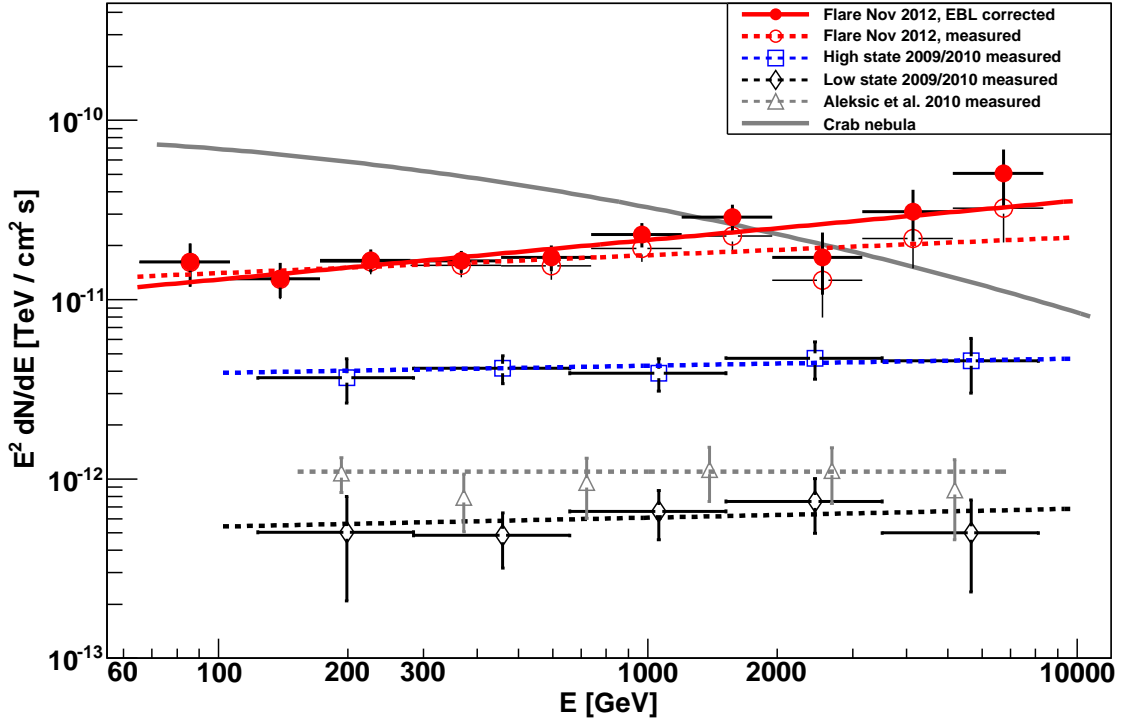} 
      \caption{
 Average spectral energy distributions during the flare (red) along with previous measurements of IC~310 as observed by MAGIC.  
 We show the results from the high (blue, open squares) and low (black, open markers) states reported in \cite{aleksic13} 
 and the average results (gray triangles) from \cite{aleksic10} for the whole period. 
 The dashed lines show power-law fits to the measured spectra, and the solid line with filled circles depicts the spectrum corrected for absorption in the extragalactic background light. 
 As a reference, the spectral power-law fit of the Crab Nebula observations from \cite{aleksic12} is shown (gray, solid line). 
 Vertical error bars show 1 standard deviation statistical uncertainity.
 Note that due to unfolding procedure spectral points are correlated.
 Horizontal error bars show the energy binning.}
               \label{SED_Flare_HighLowState}%
     \end{figure}

 \begin{figure}
    \centering
       \includegraphics[width=13cm]{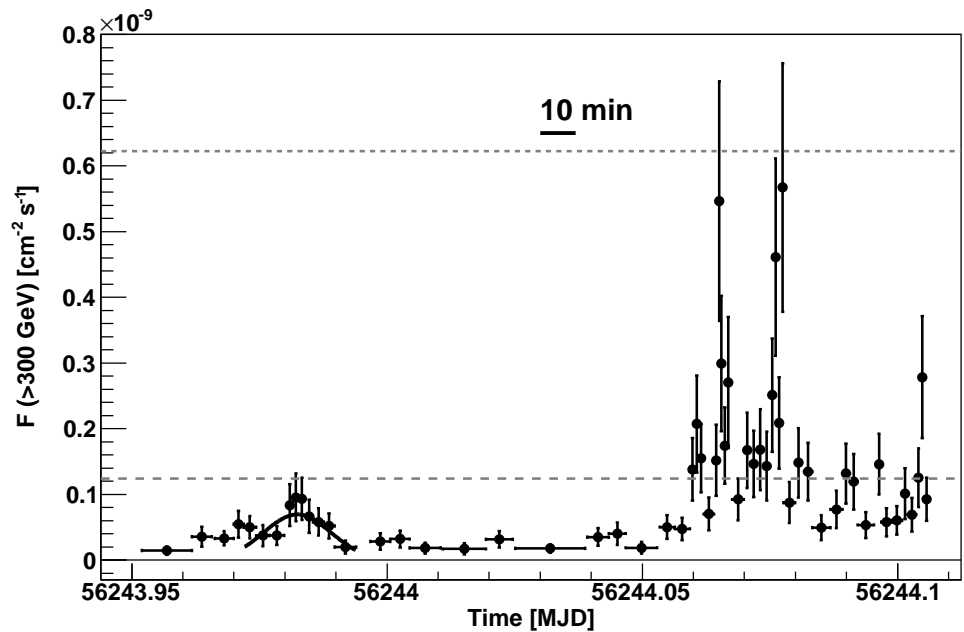}
      \caption{Light curve of IC~310 observed with the MAGIC telescopes in the night of November 12/13$^{\mathrm{th}}$, 2012, above 300\,GeV. 
     As a flux reference, the two gray lines indicate levels of $1$ and $5$ times the flux level of the Crab Nebula, respectively.
 The precursor flare (MJD 56243.972--56243.994) has been fitted with a Gaussian distribution.
 Vertical error bars show 1 standard deviation statistical uncertainity.
 Horizontal error bars show the bin widths.}
               \label{Lightcurve}%
     \end{figure}

 \begin{figure}
    \centering
       \includegraphics[width=10cm]{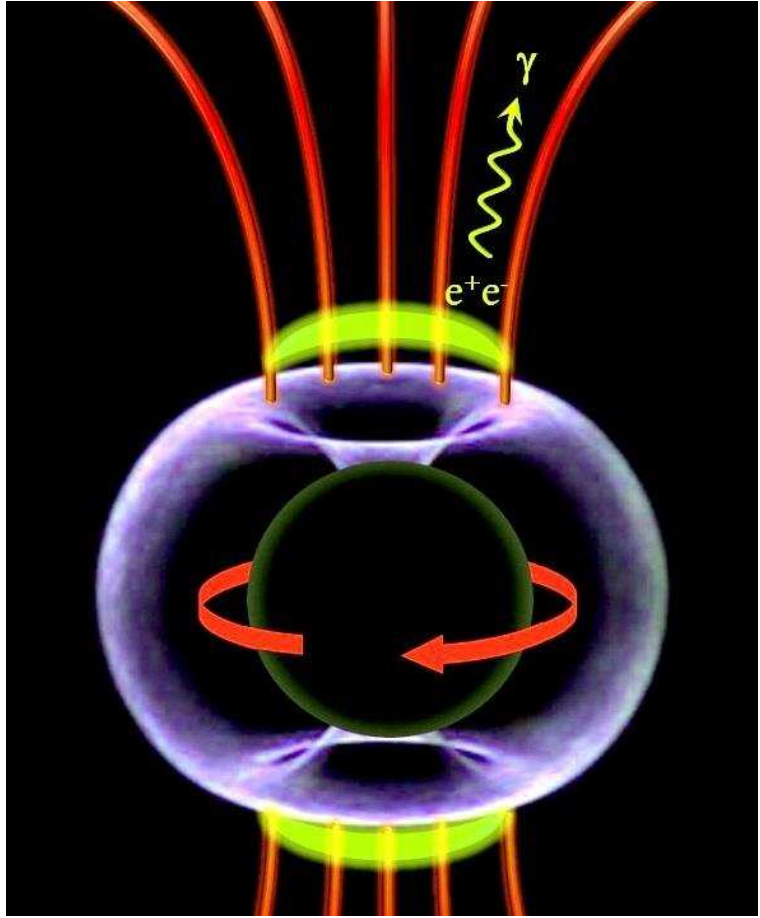}
      \caption{Scenario for the magnetospheric origin of the gamma-rays: 
A maximally rotating black hole with event horizon $r_{\rm g}$ (black sphere)
accretes plasma from the center of the galaxy IC~310.  In the apple-shaped ergosphere (blue)
extending to $2 r_{\rm g}$ in the equatorial plane, 
Poynting flux is generated by the frame-dragging effect.
The rotation of the black hole induces a charge-separated magnetosphere (red)
with polar vacuum gap regions (yellow). In the gaps, the electric field of the
magnetosphere has a component  parallel to the magnetic field accelerating
particles to ultra-relativistic energies.   Inverse-Compton scattering and copious
pair production due to interactions with low-energy thermal
photons from the plasma accreted by the black hole leads to the observed gamma rays.}
               \label{gap}%
     \end{figure}

\clearpage

\section*{Affiliations:}
 \normalsize{$^{1}$ IFAE, Campus UAB, E-08193 Bellaterra, Spain,}\\
 \normalsize{$^{2}$ Universit\`a di Udine, and INFN Trieste, I-33100 Udine, Italy,}\\
 \normalsize{$^{3}$ INAF National Institute for Astrophysics, I-00136 Rome, Italy,}\\ 
 \normalsize{$^{4}$ Universit\`a  di Siena, and INFN Pisa, I-53100 Siena, Italy,}\\ 
 \normalsize{$^{5}$ Croatian MAGIC Consortium, Rudjer Boskovic Institute, University of Rijeka and University of Split, HR-10000 Zagreb, Croatia,}\\ 
 \normalsize{$^{6}$ Max-Planck-Institut f\"ur Physik, D-80805 M\"unchen, Germany,}\\ 
 \normalsize{$^{7}$ Universidad Complutense, E-28040 Madrid, Spain,}\\ 
 \normalsize{$^{8}$ Inst. de Astrof\'isica de Canarias, E-38200 La Laguna, Tenerife, Spain,}\\ 
 \normalsize{$^{9}$ University of \L\'od\'z, PL-90236 Lodz, Poland,}\\ 
 \normalsize{$^{10}$ Deutsches Elektronen-Synchrotron (DESY), D-15738 Zeuthen, Germany,}\\ 
 \normalsize{$^{11}$ ETH Zurich, CH-8093 Zurich, Switzerland,}\\ 
 \normalsize{$^{12}$ Universit\"at W\"urzburg, D-97074 W\"urzburg, Germany,}\\ 
 \normalsize{$^{13}$ Centro de Investigaciones Energ\'eticas, Medioambientales y Tecnol\'ogicas, E-28040 Madrid, Spain,}\\ 
 \normalsize{$^{14}$ Institute of Space Sciences, E-08193 Barcelona, Spain,}\\ 
 \normalsize{$^{15}$ Universit\`a di Padova and INFN, I-35131 Padova, Italy,}\\ 
 \normalsize{$^{16}$ Technische Universit\"at Dortmund, D-44221 Dortmund, Germany,}\\ 
 \normalsize{$^{17}$ Unitat de F\'isica de les Radiacions, Departament de F\'isica, and CERES-IEEC, Universitat Aut\`onoma de Barcelona, E-08193 Bellaterra, Spain,}\\ 
 \normalsize{$^{18}$ Universitat de Barcelona, ICC, IEEC-UB, E-08028 Barcelona, Spain,}\\ 
 \normalsize{$^{19}$ Japanese MAGIC Consortium, Division of Physics and Astronomy, Kyoto University, Japan,}\\ 
 \normalsize{$^{20}$ Finnish MAGIC Consortium, Tuorla Observatory, University of Turku and Department of Physics, University of Oulu, Finland,}\\ 
 \normalsize{$^{21}$ Inst. for Nucl. Research and Nucl. Energy, BG-1784 Sofia, Bulgaria,}\\ 
 \normalsize{$^{22}$ Universit\`a di Pisa, and INFN Pisa, I-56126 Pisa, Italy,}\\ 
 \normalsize{$^{23}$ ICREA and Institute of Space Sciences, E-08193 Barcelona, Spain,}\\
 \normalsize{$^{24}$ Universit\`a dell'Insubria and INFN Milano Bicocca, Como, I-22100 Como, Italy,}\\ 
 \normalsize{$^{25}$ now at: NASA Goddard Space Flight Center, Greenbelt, MD 20771, USA and Department of Physics and Department 
of Astronomy, University of Maryland, College Park, MD 20742, USA,}\\ 
 \normalsize{$^{26}$ now at Ecole polytechnique f\'ed\'erale de Lausanne (EPFL), Lausanne, Switzerland,}\\ 
 \normalsize{$^{27}$ Now at Institut f\"ur Astro- und Teilchenphysik, Leopold-Franzens- Universit\"at Innsbruck, A-6020 Innsbruck, Austria,}\\ 
 \normalsize{$^{28}$ now at Finnish Centre for Astronomy with ESO (FINCA), Turku, Finland,}\\ 
 \normalsize{$^{29}$ now at Astrophysics Science Division, Bhabha Atomic Research Centre, Mumbai 400085, India,}\\ 
 \normalsize{$^{30}$ also at INAF-Trieste,}\\ 
 \normalsize{$^{31}$ now at School of Chemistry \& Physics, University of Adelaide, Adelaide 5005, Australia,}\\ 
\normalsize{$^{32}$ Dr. Remeis-Sternwarte Bamberg, Astronomisches Institut der Universit\"at Erlangen-N\"urnberg, ECAP, D-96049 Bamberg, Germany,\\
\normalsize{$^{33}$ Max-Planck-Institut f\"ur Radioastronomie, D-53121 Bonn, German,\\
\normalsize{$^{34}$ Observatori Astron\`omic, Universitat de Val\`encia, E-46980 Paterna, Val\`encia, Spain,\\
\normalsize{$^{35}$ Departament d'Astronomia i Astrof\'{\i}sica, Universitat de Val\`encia, E-46100 Burjassot, Val\`encia, Spain \\
\normalsize{$^{\dag}$ deceased

\newpage

\begin{center}
 \Large{Supplementary Materials for}\\
 \vspace*{0.5cm}\large{Black hole lightning due to particle acceleration at subhorizon scales}\\
 \vspace*{0.5cm}
 \normalsize{$^{\dag}$correspondence to: deisenacher@astro.uni-wuerzburg.de (D. Eisenacher),}\\
 \normalsize{jsitarek@ifae.es (J. Sitarek), mannheim@astro.uni-wuerzburg.de (K. Mannheim)}
\\
\end{center}

\vspace*{1.5cm}
\noindent\textbf{This PDF file includes:}\\
\hspace*{0.5cm} Materials and Methods\\
\hspace*{0.5cm} Fig. S1 to S5\\
\hspace*{0.5cm} Tables S1 to S2\\
\hspace*{0.5cm} References (42-69)

\newpage

 \noindent \textbf{\Large{Materials and Methods}}\\
 \vspace*{-0.6cm}

\section{Comments on IC\,310}

The nearby lenticular (S0, $z=0.0189$) galaxy IC\,310 at a distance of 81\,Mpc exhibits an active galactic 
nucleus (AGN). 
This object has been detected at high energies (HE) above 30\,GeV with 
\textit{Fermi}/LAT (\textit{26}) as well as at very high energies (VHE) above 260\,GeV 
(\textit{27, 28}). \\
In the past, the object was thought to be a head-tail radio galaxy \cite{ryle68, sijbring, miley80}. 
Those are typically found in clusters of galaxies and they are characterized by their extended 
jets which are pointing away from the center of the cluster determined by the galaxy's motion 
through the intra-cluster medium (ICM).
However, using the Very-Long-Baseline Interferometry (VLBI) technique, the large-scale jet was 
found to follow the direction of its parsec-scale one-sided jet to within about $10^\circ$ (\textit{32}).
Hence, there is no indication of the interaction
of the VLBI jet with the ICM that would determine the 
direction of the tail. 
Instead, the inner jet appears to be blazar-like with a missing counter jet due to relativistically boosted emission.
Further indications for transitional behavior between a radio galaxy and a blazar was 
found in IC\,310 in various energy ranges \cite{rector}, e.g., weak optical emission lines 
similar to Fanaroff-Riley I (FR~I) radio galaxies \cite{owen}, and a non-thermal point-like 
emission in the X-ray regime \cite{schwarz, rhee, sato}. In addition, a faint halo of X-ray radiation
in the direction of kpc radio jet has been found \cite{dunn}. 

\subsection
{Mass of the IC\,310 black hole.}
The mass, $M$,  of the black hole of IC\,310 can be inferred from its relation with the velocity dispersion, $\sigma$, of the host galaxy (\textit{29}), namely: 
\begin{equation}
 \mathrm{log}\left(\frac{M}{\mathrm{M}_{\odot}}\right)=\alpha+\beta \mathrm{log}\left(\frac{\sigma}{200\,\mathrm{km\,s}^{-1}}\right)
\end{equation}
The velocity dispersion of $\sigma=(229.6\pm5.9)$\,\,km\,s$^{-1}$ has been measured for IC\,310 (\textit{30, 51}).
Applying $\alpha=(8.12\pm0.08)$ and $\beta=(4.24\pm0.41)$ 
(\textit{29}) yields a mass estimate of $M_{\mathrm{BH}}= (2.4\pm0.5)\times\,10^{8}$\,M$_{\odot}$. 
For early-type galaxies such as those with Hubble-type S0 $\alpha=(8.22\pm0.07)$ and $\beta=(3.86\pm0.38)$ were found. 
This results in a slightly higher mass of $M_{\mathrm{BH}}= (2.8\pm0.6)\times\,10^{8}$\,M$_{\odot}$.
The uncertainty has to however take into account also an intrinsic scatter of $\log (M/\mathrm{M}_{\odot})$, which is $\epsilon_0=0.44\pm0.06$ or $\epsilon_0=0.35\pm0.03$ for the two above estimations respectively. 
We conclude that the mass of IC\,310 can be estimated from the $M-\sigma$ relation with precision of $M=(3^{+4}_{-2}) \times10^{8}M_\odot$.

A second method to estimate the mass of a black hole is based on the fundamental plane of black hole activity (\textit{31}).
This fundamental plane was inferred by studying the X-ray $L_{\mathrm{X}}$ in the 2--10\,keV range and radio core $L_{\mathrm{R}}$  
luminosities at 5\,GHz of stellar mass and supermassive black holes.
The following \textit{fundamental} correlation between these luminosities and the mass of the black hole was found:
\begin{equation}
 \mathrm{log}\,L_{\mathrm{R}}=\left(0.60^{+0.11}_{-0.11}\right)\mathrm{log}\,L_{\mathrm{X}}+\left(0.78^{+0.11}_{-0.09}\right)\mathrm{log}\,M_{\mathrm{BH}}+7.33^{+4.05}_{-4.07},
\end{equation}
with a large scatter of $0.88$.
To calculate the mass for IC\,310, the EVN observation at 5\,GHz from October 29, 2012 is used. The total flux density 
was measured to be $S_{5\,\mathrm{GHz}}=0.109$\,Jy with an uncertainty of 10\%.  
This results in a radio luminosity of $L_{\mathrm{R}}=(4.3\pm0.4)\times10^{39}$\,erg\,s$^{-1}$.  
The X-ray luminosity was calculated from the \textit{Swift}-XRT observation on November 14, 2012 \cite{eisenacher13}
$F_{2-10\,\mathrm{keV}}=(6.4\pm0.4)\times10^{-12}$\,erg\,cm$^{-2}$\,s$^{-1}$, hence
$L_{\mathrm{X}, 2-10\,\mathrm{keV}}=(5.0\pm0.3)\times10^{42}$\,erg\,s$^{-1}$.
This second method nominally yields $M\sim 4 \times10^8$\,M$_{\odot}$, which is consistent with the value obtained from $M-\sigma$ relation.
However the large scatter in the fundamental plane does not allow us to further improve the mass estimate.

\subsection
{Constraining the angle to the line-of-sight.}
Interferometric observations in the radio band constrain the angle between the jet axis of an AGN and the line of sight by
measuring the ratio between the brightness of the jet and its counter-jet assuming relativistic
Doppler boosting. In case of a non-detection of the counter-jet ($<3\sigma$ in the image) only an upper limit for the angle can be estimated. 
Using results from observations with the European VLBI Network (EVN) 
the angle can be constrained to be less than $\lesssim20^{\circ}$ (see Section EVN). \\
Additionally, the extension of the projected one-sided kpc radio jet of $\sim350$\,kpc measured 
at a wavelength of 49\,cm \cite{sijbring} yields an estimate of a lower limit for the angle.
De-projecting the jet using the upper limit quoted above would results in a lower limit of the jet length of $\sim$1\,Mpc. 
Radio galaxies typically show jets extending up to 150\,kpc-300\,kpc 
\cite{neeser95}. The maximal length of radio jets has been measured to be a few Mpcs which corresponds to an angle of $\sim5-10^{\circ}$ in 
the case of IC\,310. Smaller angles would rapidly increase the de-projected length of the jet to values far above the maximum of the 
distribution of the jet lengths. 
Such extremely elongated jets are inconsistent with the advance speed of jets into the intergalactic medium, 
the masses, and lifetimes of AGNs. 
Adopting the typical advance speed of 0.01\,c \cite{parma} and the AGN lifetime of 
$10^8$ years \cite{sijbring}, the length of a jet becomes 
$\sim300$\,kpc. The latter is known from the ratio between the space density of AGN and 
their host galaxies or can be obtained from the calculation of the electron ageing along the kpc jet
using observations of a cooling break in their radio spectra. 
For IC\,310, the lifetime of electrons far away from the radio core has been determined in this way to be
$2-2.5\times10^8$\,years \cite{feretti98}
corresponding to a jet length of $\sim800$\,kpc and an orientation angle of 26$^{\circ}$.
Accreting at an Eddington rate of several solar mass per year, the supermassive black hole grows to a 
mass of the order of $10^8$\,M$_\odot$ in this time span, consistent with the mass of 
the black hole given above.\\
Since VLBI jet components often show position angles within the
range of the jet opening angle \cite{lister13}, the high-energy
radiation emitting jet base could be oriented in a different direction.
Note that a bend in the direction to achieve a Doppler-favorism angle $\sim\Gamma_{\mathrm{j}}^{-1}$ is 
barely likely for arbitrary orientations of jets in the Perseus cluster finding one exactly 
pointing towards us.
Assuming a semi-opening angle of $6^{\circ}$, which corresponds to a Lorentz factor of $\Gamma_{\mathrm{j}}=10$,
the probability that the jet will randomly point in a certain direction is 
$1-\cos(6^\circ)$ = 0.5\%\footnote{Notice that there are two jets in a sphere of $4\pi$. This estimate does not 
take into account the observational biases due to a higher detection probability in the TeV range for AGN with smaller
angle between the jet axis and the line of sight.}. 
Furthermore, a bend not affecting the position angle 
between the projected pc and kpc-scale jet 
would multiply yet another 
chance probability of 11\%.

\subsection
{The jet power of IC\,310.}

As some models for the production of the flaring VHE radiation require the parameter
jet power we will estimate it inferred from radio measurements.
Estimating the jet power from the minimum energy assumption and the observed synchrotron cooling break
in the radio yields $L_{\mathrm{jet}}= 2\times10^{42}$ erg\,s$^{-1}$ \cite{sijbring}.  Taking into account protons with the cosmic
ray proton-to-electron energy density ratio of 50 \cite{blandford87}, the minimum energy increases by $50^{4/7}$, and the
corresponding jet luminosity attains the value of $\sim2\times10^{43}$\,erg\,s$^{-1}$.

\section{Observation and analysis of the MAGIC and EVN data}

\paragraph*{MAGIC.} The observations of the flare of IC\,310 were carried out with the MAGIC telescopes, two
Imaging Atmospheric Cherenkov Telescopes located on the island of La Palma (28.4$^{\circ}$N, 17.54$^{\circ}$W) at an altitude of 2200\,m above sea level.
Each of the telescopes consists of a mirror dish of 17\,m diameter and is equipped with a fast imaging 
camera with a field of view of 3.5$^{\circ}$.
Recently, between 2011 and 2012 the MAGIC-I camera has been replaced by a close copy of the MAGIC-II 
one and now consists of 1039 pixels with a diameter of 0.1$^{\circ}$ each \cite{mazin13}. 
At the same time also the size of the trigger region was enlarged. 
In addition, the readout system of both telescopes has been upgraded to a digitizing system based on 
the Domino-Ring-Sampler version 4 chip \cite{sitarek13b}.
Shortly after the final commissioning of the upgraded system on November 12/13, 2012 (MJD 56243.95--56244.11)
an exceptional flare of IC\,310 was detected during observations of the Perseus cluster. 
In this night the observation of 3.7\,h in total was performed in the so-called wobble mode. 
In this mode, the telescopes are pointing alternating every 20 minutes between four sky positions, 
each with an offset of 0.4$^{\circ}$ to the object of interest. 
In case of the Perseus cluster, the particular wobble pointings were selected to optimize the exposure 
of the two VHE objects (namely NGC\,1275 \cite{aleksic12a} and IC\,310) already detected in this field-of-view by MAGIC. 
The wobble center of those observations was chosen to be the center of the cluster, i.e., NGC\,1275. 
Two of the wobble positions have an offset of 0.4$^{\circ}$ with respect to IC\,310 and the 
remaining ones are at a distance $0.94^{\circ}$ away from the object.

The signal extraction and calibration of the data, the image parametrization, the direction and energy reconstruction 
as well as the gamma-hadron separation was applied with the standard analysis software MARS as 
described in \cite{zanin13}. 
The performance of the MAGIC telescopes after the upgrade has been presented in \cite{sitarek13}.

For the calculation of the gamma-ray flux and the spectrum the effective area of the detector is needed 
which can be obtained from Monte Carlo simulations of gamma-ray events.
However, the effective area strongly depends on the distance of object to the camera center. 
In particular, the collection area above 300\,GeV is roughly twice smaller for the offset of 
0.94$^{\circ}$ than for 0.4$^{\circ}$. 
Since during the observation of the Perseus cluster wobble pointings with these two different offsets 
to IC\,310 were used, also two sets of Monte Carlo simulations 
with the corresponding wobble offsets have been produced.

In the night of the flare a clear signal (see Fig.~\ref{figS1}) of 507 gamma-like events
in the region around IC\,310 in excess of the background with
a significance of 32\,$\sigma$ above 300\,GeV could be reconstructed. 

Due to still limited statistics of the $\gamma$-ray and background events and
 a very rapid variability behavior, the classical approach for the calculation of light curves 
in gamma-ray astronomy which is based on the fixed width of the time bins is not optimal. 
We used instead a method similar to the one commonly used for data of X-ray observatories for the 
computing of energy spectra. We first identify all periods in the data during which the telescopes were not operational 
(in particular $\lesssim 1$\,min gaps every 20\,min when the telescope is slewing and 
reconfiguring for the next data run.) 
Afterwards, we bin the remaining time periods based on a fixed number (in this case 9) of ON events (see Fig.~\ref{figS1})
per bin. We estimate the number of background events in each time bin from four
off-source regions at the same distance from the camera center.
As the signal to background ratio above 300\,GeV is much larger than 1 (see Fig.~\ref{figS1}) this assures that the 
precision of individual points in the light curve is close to $3\sigma$.

In addition, we performed a toy Monte Carlo study comparing light curves obtained with both methods (see Fig.~\ref{figS2}). 
If the number of events per bin is low ($N_{\mathrm{ex}}\lesssim 10$) a downfluctuation of a number of events in a 
fixed width time bin biases the standard error estimated as $\sqrt{N_{\mathrm{ex}}}$ to lower values. 
This results in the low tail in the distribution of the residual, bias in the mean and the RMS of the 
resulting distribution (in the examples presented here Mean\,$=-0.1$ to $-0.2$, RMS\,$=1.05$ to $1.15$). 
The characteristic, discrete shape of the distribution in this case is due to integer, Poissonian 
statistics of the number of events in a time bin.
On the other hand the residual distribution in the case of the fixed number of ON events method is 
continuous and does not produce any net bias (Mean = 0, RMS = 1). 
Both classical, and the new light curve method show an asymmetric behavior of the residual distribution 
(notice the logarithmic scale in Y axis of the bottom panels). 
This however cannot be easily corrected without introducing an assumption on the unknown shape of the 
light curve.

\paragraph*{VHE light curve.} 

The light curve above 300\,GeV measured in the time range of MJD 56243.95--56244.11 is shown Fig.~\ref{figS3}. 
Here, we present the light curve with fixed time bins of 3 minutes as well as binned with 9 ON 
events per bin. Both results are in good agreement. 
However, the light curve with fixed ON events is more sensitive for identifying 
ultra-fast ($<1$ minute) flux variations.    
The mean flux above 300\,GeV during this period is 
$\Phi_{\mathrm{mean}}=(6.08 \pm 0.29)\times10^{-11}$\,cm$^{-2}$s$^{-1}$. This is four times higher than the
high state flux of $(1.60 \pm 0.17)\times10^{-11}$\,cm$^{-2}$s$^{-1}$ reported in (\textit{28}).

Fitting the fixed ON events light curve in the full time range with a constant reveals a 
$\chi^2/\mathrm{N.d.o.f}$ of 199/58 and a probability of $2.6\times10^{-17}$ to be constant.
In the time range of 56244.05--56244.15 the flux seems to be higher. Here, the constant fit gives 
a $\chi^2/\mathrm{N.d.o.f}$  of 80/35 and a probability of $2.2\times10^{-5}$.

In order to find the time scale in which the flux is doubled/halved we applied the following fit 
to individual substructures in the light curve:
\begin{equation}
 F(t) = F(t_0)*e^{-\frac{t-t_0}{\tau}}
\end{equation}
where $F(t)$ and $F(t_0)$ are the fluxes at the time $t$ and $t_0$, respectively.
The doubling time $\tau_{\mathrm{D}}$ is then given by $\tau\times \mathrm{ln}2$. Some exemplary results of those fits 
are shown in Fig.~\ref{figS4} and in Table~\ref{tabS1}. Notice that the choice of the fit range may affects the result of the doubling time significantly.

The doubling times obtained from different methods are ranging from $\sim1-10$ minutes (see also Fig.~\ref{figS3}). 
As the pre-flare shows a similar time scale for the rising and the
decaying period we fitted it
with a Gaussian function (Fig.~\ref{figS4} left panel). The standard deviation
of this Gaussian function obtained from the fit is $(9.5\pm1.9)$\,min.\\

We use the rapidly rising part of the 1$^{\mathrm{st}}$ big flare (MJD 56244.062--56244.0652) in order to compute the
conservative, slowest doubling time, $\tau_\mathrm{D}$, which is still
consistent with the MAGIC data.
We fit the light curve with a set of exponential functions, each time
fixing $\tau_\mathrm{D}$ to a different value and computing the
corresponding fit probability.
We obtain that $4.9\,\mathrm{[min]}$ is the largest value of
$\tau_\mathrm{D}$, which can still marginally fit the data with
probability $>5\%$ (see the dashed line in the right panel of Fig.~\ref{figS4}).
Note that the corresponding time scale in the frame of reference of IC 310 will be slightly shorter: $4.9/(1+z)\,\mathrm{[min]} = 4.8\,\mathrm{[min]}$.

\paragraph*{VHE spectrum.} 

The observed spectrum can be described by a simple power law:
\begin{equation}
\frac{\mathrm{d}F}{\mathrm{d}E}=f_0\times\left(\frac{E}{1\mathrm{TeV}}\right)^{-\Gamma}.
\end{equation}
Results of the flux normalization at 1\,TeV $f_0$ and, the photon index $\Gamma$ are given in 
Table~\ref{tabS2}.
In addition, the results of the intrinsic spectra, i.e. corrected for the absorption through the 
extragalactic background light according to \cite{dominguez11} are given. 
Due to the proximity of IC\,310 the effect is only minor and causes a change of 20\% in the 
flux normalization and $\sim0.1$ of the index as already reported in (\textit{28}). 

As part of the observation was carried out with a higher then usual
offset angle from the camera center the systematic error on the flux
normalization is slightly larger (12\%) then reported in (\textit{25}). The error
of the energy scale is 15\%.

In (\textit{28}) it has been reported that the VHE spectrum of IC\,310 does not change when the 
object seems to be in different emission states.
To evaluate the spectral behavior during the flare the so-called hardness ratio has 
been investigated as presented in Fig.~\ref{figS5}.
We defined the hardness ratio as the ratio of integral flux above 1\,TeV to the one in the energy 
range 300--1000\,GeV. As the distribution presented in the left panel 
of Fig.~\ref{figS5} is compatible with a linear correlation and the hardness ratio is constant with a probability of 0.16 
(Fig.~\ref{figS5}, right) no significant spectral variability 
could be detected but a hint for a spectral hardening with increasing flux is present.

By calculation of individual spectra of several time ranges during the flare (according to different flux levels) 
 allowed an other possibility for searching for spectral variability. From this study 
only a hint for spectral hardening with increasing flux can be found.

\paragraph*{EVN.} 

IC\,310 has been observed with the European VLBI Network (EVN) at 1.7, 5.0, 8.4 and 22.2\,GHz between 2012-10-21 and 2012-11-07.
The data were amplitude and phase calibrated using standard procedures with the Astronomical Image Processing System (\textsc{AIPS}, 
\cite{Greisen2003}) and imaged and self-calibrated using \textsc{DIFMAP} \cite{Shepherd1994}. 

Here, we present the image (see Fig.~2 of main paper) with the highest dynamic range obtained from the observation at 5.0 GHz from 2012-10-29. 
The participating telescopes were:
 Effelsberg, Westerbork, Jodrell Bank, Onsala, Medicina, Noto, Torun, Yebes, Zelenchukskaya, Badary, Urumqi, and Shanghai.
Due to technical problems the calibration information for Jodrell Bank, 
Zelenchukskaya and Badary was incomplete. Hence, we produced an image in \textsc{DIFMAP} without these telescopes first. 
The resulting image model was used to determine a constant amplitude correction factor for each telescope, with which the amplitude 
calibration in \textsc{AIPS} and imaging in \textsc{DIFMAP} was repeated for the complete array. 

The final image has a peak flux density of 77\,mJy/beam and a 1$\sigma$ noise level of 0.027\,mJy/beam.
The restoring beam has 
a major and minor axis of $4.97\times1.24$\,mas$^{2}$ with the major axis at a position angle of $-8.5^{\circ}$. 
It contains a total flux density of $S_\mathrm{total}=109\mathrm{\,mJy}$, which we conservatively assume to be 
accurate to 10\%. The dynamic range $DR$ of the image, i.e., the ratio of the peak flux density and three times the noise level in the image is $DR\approx 950$. 

The angle $\theta$ of the radio jet to the line-of-sight can be determined from Doppler boosting arguments for a given jet speed $\beta$ and spectral 
index $\alpha$ by considering the ratio $R$ of the flux density in the jet and counter-jet:
\begin{equation}
 R=\left(\frac{1+\beta\cos\theta}{1-\beta\cos\theta}\right)^{2-\alpha}.
\end{equation}
Following (\textit{32}) we use the $DR$ as an upper limit for the detection of a counter-jet. This gives us an upper limit of $\theta$:
\begin{equation}
 \theta < \mathrm{arc\,cos} \left(\frac{DR^{1/(2-\alpha)}-1}{DR^{1/(2-\alpha)}+1}\right).
\end{equation}
 
Substituting $DR$ in Eq.\,(6), assuming a flat spectral index of $\alpha=0$ and
$\beta\rightarrow1$, we obtain an upper limit for the angle between the jet and the line-of-sight of $\lesssim 20^{\circ}$.

\section{Gamma-ray production models for the flare}

In this section we review possible classes of gamma-ray
emission models and evaluate their applicability to the
case of the flare of IC\,310. The important observational behaviour that needs to be explained
by the models are: (i) frequent VHE activity of a low luminosity radio galaxy, i.e.,
almost no Doppler boosting, (ii) minute-scale variability on top of a VHE flare, (iii) a hard, simple power-law VHE photon spectrum 
up to energies of 10\,TeV, and
(iv) no significant spectral variability of the VHE emission during different flux states.     

\subsection{Cloud/Star-jet interaction models} 

The interaction of jets with gas clouds was originally proposed as a mechanism for
mass entrainment and flux variability in the radio band (\textit{1}). 
Gas clouds are known to exist in AGN due to their optical emission lines,
and they could originate, e.g., from the stellar winds of massive stars with high mass losses.
A number of studies deals with their role in producing gamma-ray variability.

\paragraph*{Cloud-in-a-jet model.}
 In (\textit{35, 36}) a model is presented that tries to explain the observation of a VHE flare 
from the radio galaxy M87 in April 2010. 
This event was modeled by a dense gas cloud with mass of $>10^{29}$\,g, e.g., originating from an 
atmosphere of a red giant (solar-mass-type star)
that travels through the jet of M87, blows up and forms an accelerated gas cloud which is heated by 
the jet pressure. The authors assume a strongly magnetized jet base in which electrons may not reach 
TeV emitting energies.
Hence, the model applied here is based on emission produced by proton-proton ($pp$) interactions of 
accelerated protons located at the jet-cloud interface. It explains day-scale variability as well as a 
hard VHE spectrum
without specifically describing the underlying acceleration process of the particles. Both, magnetic reconnection as 
well as Fermi I and II processes are possible in this model. 
Such an event produces one single peak in the light curve which depends on the size of the expanding 
cloud. Several peaks can be achieved if the star envelop splits into fragments. In general, this could 
produce variability on even shorter time scales, however,
a very large relativistic beaming factor is necessary in order to reach the observed luminosity
which is inconsistent with the non-relativistic cloud velocity. 
Therefore, this model does not seem feasible in the case of the minute-scale 
variability of IC\,310.

\paragraph*{Star-in-a-jet model.} The model presented by (\textit{34}) is based on an acceleration of high 
energy particles in the shocks occuring between the jet and wind of a star immersed in the jet's plasma. 
The authors claim that at any given time a few tens of massive stars can be found inside the inner 
parsec of a typical AGN jet. 
Strong winds of WR or young OB stars would then collide with the 
jet causing a double shock structure.
On those shocks electrons can be accelerated along the shock structure.
The relativistic electrons will then scatter the thermal radiation of an OB star producing very high 
energy gamma-rays. 
The model was originally used for explanation of a day time scale variability in the blazar Mrk\,421. \\
A similar scenario can be considered in the case of the IC\,310 flares as the  star-jet shocks can be 
directed at relatively large angles to the jet direction.
Let's consider a typical OB star with a radius of $R_{s}=10^{12}\,$cm and mass loss rate of 
$10^{-6}M_\odot /\mathrm{yr}$ and wind speed of $v_{w}=10^{3}$\,km/s.
The star is moving through the jet with velocity of $10^{4}$\,km/s at the distance of 0.01\,pc from its base.
Using Eq. (4) in (\textit{34}) we can estimate the radius of the shock as $4\times10^{13}\,$cm for 
the kinetic power of the jet of $2\times 10^{42}\,$erg\,s$^{-1}$.
The time scale of the individual flares seen from IC\,310 would be then a result of an instability of 
the shock direction caused by the inhomogeneous wind of the massive star.
These irregularities would have to be of the order of 3\% of the stellar dimension for the time scale for the change 
of shock direction, $\sim 0.03\cdot R_{s}/v_{w}=5$\,min to be comparable to those seen from 
the flare.
The maximum energies to which particles can be accelerated in such a model depend on the acceleration 
coefficient and the star's magnetic field at the distance of the shock.
The typical surface magnetic field of OB stars is of the order of $1\,$kG \cite{igoshev11}. 
Assuming dipole structure, the magnetic field at the shock should be of the order of $0.01\,$G.
With acceleration coefficients of $1.6\times10^{-4}-0.04$ the electrons might be accelerated up to $\sim20-100$\,TeV.
An unbroken gamma-ray spectrum with an index close to $-2$ extending up to 10\,TeV, as observed 
from IC\,310, can then be produced due to efficient cooling of electrons in the Thomson and 
Klein-Nishina regimes by scattering radiation from massive star.\\
We also evaluate if such a model can explain the isotropic luminosity seen from IC\,310 during the flare, 
$\sim 2\times10^{44}$erg\,s$^{-1}$. 
In the above considered model, the total luminosity extracted from the jet is independent from the 
jet power (less powerful jets will allow the stellar wind to expand farther, thus covering with a 
shock a larger fraction of the jet), therefore relatively small jet power of the IC\,310 is not 
problematic.
However the fact that only a small fraction of the jet (around the star) can produce the emission is 
limiting the allowed luminosity to $\sim\,5\,\times\,10^{38}\,\mathrm{erg}\,\mathrm{s}^{-1}$, which is significantly below the 
isotropic VHE luminosity seen during the IC\,310 flares.
In order to produce such flux, the emission has to be beamed into a cone with the total opening angle 
not larger than $0.4^\circ$.
Summarizing such a model can explain the spectrum and might explain the variability time scales of the flare, 
however requires rather extreme beaming to explain the flux level.

\subsection{Jets-in-a-jet model}

One consideration in explaining rapid variability of TeV emitting AGNs and avoiding
absorption of the TeV photons by $\gamma\gamma$ pair production has been developed by dividing 
the relativistic jet ($\Gamma_{\mathrm{j}}$) into several subregions with $\Gamma_{\mathrm{mj}}$ that 
move relatively to the main jet, e.g., (\textit{21, 33}).
This can solve the problem if the variability timescale is shorter than gravitational timescale 
$\Delta t_{\mathrm{BH}}\equiv G_{\mathrm N}M/c^3$ of the black hole. The jets-in-a-jet approach insures the size of the 
emitting region $r$ not being smaller than $r_\mathrm{g}$
because only small fraction of $r$ that is beamed into a narrow cone is responsible for the observed 
short variability time.
In general, the 'emitted' Lorentz factor obtained in those models is 
$\Gamma_{\mathrm{em}}\propto\Gamma_{\mathrm{j}}\Gamma_{\mathrm{mj}}$, where $\Gamma_{\mathrm{j}}$ 
denotes the Lorentz factor of the jet and $\Gamma_{\mathrm{mj}}$ the Lorentz factor of 
the mini-jets, respectively. Assuming $\Gamma_\mathrm{j}=\Gamma_{\mathrm{mj}}=10$ one gets a total 
Lorentz factor of $\Gamma_{\mathrm{em}}=100$ for blazars as required in (\textit{15}) in order 
to avoid absorption of the TeV photons.

Such a model has been proposed for daily-scale variability of M87 (\textit{33}), 
and for the minute-time flaring blazars as Mrk\,501 or PKS\,2155$-$304 (\textit{21}) by assuming 
blobs of energetic particles that are moving relatively to the direction of the jet. Those minijets can be 
produced through dissipation of magnetic energy in a Poynting-flux-dominated jet. Such magnetic 
reconnection events might be a result, e.g., 
of kink instabilities. For M87 the authors note that flares on time scales of 2-3 hours can still 
be explained by their model.
Below we will adapt this approach to the observational constraints given by IC\,310. We fix in the 
following basic data:
$L_{\rm rad}\sim L_{\gamma}\simeq 2\times10^{44}$ erg s$^{-1}$, 
and $t_{\rm var}=4.8$\,min. 
We consider 
the following benchmark parameter values for the jet and the mini-jets:
$\Gamma_{\rm j}=10$ and $\Gamma_{\rm mj}=10$.
The parameter $\alpha$ defines the ``off-axis'' observation: 
\begin{equation}
\alpha \equiv \frac{\theta}{1/\Gamma_{\rm j}}.
\end{equation}
Assuming $\alpha=2$ to account for the difference between the narrow beaming cone and the observation angle,
 this results in a total Lorentz factor of:
\begin{equation}
\Gamma_{\rm em} \sim \frac {2 \Gamma_{\rm j} \Gamma_{\rm mj}} {\alpha^2}=50.
\end{equation}
The size of the emitting blobs can be determined as: 
$R_{\mathrm{mj}}=c \, t_{\rm var} \Gamma_{\rm em}=4.3\times 10^{14}$ cm.
Following (\textit{31}), the required jet power can be determined in the following way: 
the  observed luminosity from a mini-jet in a jet observed off-axis can be estimated to be 
(see Eq. 6 in (\textit{33}):
\begin{equation}
L_{\rm obs}=\frac{16L_{\rm on}}{\alpha^8}
\end{equation}
from which we can derive the on-axis luminosity $L_{\rm on}$.
The beaming-corrected {\rm emitted} luminosity from this mini-jet would be:
\begin{equation}
L_{\rm rad}=\frac{L_{\rm on}}{4\Gamma_{\rm em}^2},
\end{equation}
and the power required to sustain such a radiative luminosity is $L_{\rm mj}=L_{\rm rad}/\epsilon$,  
$\epsilon $ being the radiative efficiency.
For one mini jet pointing toward the observer there will be a total of $N\sim \Gamma_{\rm mj}^2$ 
active mini jets. The {\it total} power of the mini-jets would therefore be:
\begin{equation}
L_{\rm tot}=\frac{L_{\rm obs}}{\epsilon} \frac{\alpha^{12}}{256 \, \Gamma^2_{\rm j}},
\end{equation}
in which we used Eq. 4 of (\textit{33}).
For $\alpha=2$, $L_{\rm obs}=2\times 10^{44}$ erg s$^{-1}$ and assuming $\epsilon =0.1$ we finally 
obtain:
\begin{equation}
L_{\rm tot}\simeq 3\times 10^{44} \,\, {\rm erg \,s}^{-1}.
\end{equation}
If the mini-jets tap a fraction $f$ of the total jet power, the latter is:
\begin{equation}
L_{\rm j}\simeq 3\times 10^{44} f \,\, {\rm erg \,s}^{-1},
\end{equation}
which is clearly higher than the jet power inferred from radio observations. 
 
Furthermore, a VHE photon spectrum up to 10\,TeV without a break might produce problems in explaining
the observation with this model as the acceleration process must then be very efficient under the condition
of a small magnetic field and/or a small acceleration time scale assuming a fast escape of the particles. 
Following the discussion in (\textit{33}) we can also estimate the minimal variability timescale
that can still be explained by the model. Since the black hole of IC\,310 has a mass $\sim$4 times smaller than in PKS\,2155$-$304 and Mrk\,501, 
and assuming again $\alpha=2$, we can expect variability on timescales of the order of $(5\,\mathrm{min}/4)\times4=5$\,min comparable to 
 the observed timescales.

\subsection{Magnetosphere models}

The observation of short-time variability in the radio galaxy M87 has led to a renewed interest
in magnetospheric models for the gamma-ray emission, e.g., (\textit{4, 39, 40, 68}). 
Owing to the low accretion rate prevalent in radio galaxies, the magnetosphere induced by the rotating supermassive black 
hole suffers from particle starvation leading to a breakdown of the screening of electric fields. 

The jet luminosity corresponds to an accretion rate of $\dot m\sim 10^{-4}$
in units of the Eddington accretion rate, in which case
the black hole magnetosphere is just below the borderline to particle starvation, i.e.,
\begin{equation}
{n_{\mathrm{e}^\pm}\over n_{\mathrm{GJ}}}=0.02\, \dot m_{-4}^{3.5}\, m_8^{0.5}.
\end{equation}
Inevitably, for charge densities $n_{\rm e^\pm}$ well below the Goldreich-Julian density $n_{\rm GJ}$, a force-free magnetosphere cannot be 
maintained and electrostatic gaps will form that accelerate beams
of particles \cite{burns82}. Although the gap size $h$ depends sensitively on the available seed particles and the 
particle multiplicity of the electromagnetic cascades initiated by the accelerated seed particles, sub-horizon scale dimensions compatible
with the gamma-ray constraints on the size of the emission region
$h\sim0.2 r_{\rm g}$ are possible in this scenario.  
In radiatively inefficient accretion flows (RIAF),
the absorption of gamma rays due to collisions with low-energy thermal synchrotron photons  can be neglected in the TeV 
range for a broad range of parameters.
Eventually, the runaway particle production in the cascades leads to an increase of the charge carrier density that 
shortcuts the potential across the gap, terminating the flare.  
Afterwards, the particles are swept away with the jet flow, the gap can reopen giving rise to further flares.  
The potential drop across the gap 
is significantly larger than 10\,TeV and
the observed spectrum results from unsaturated electromagnetic cascading, 
implying a rather stable
spectrum with a photon index close to 1.9 \cite{mannheim93}.
The parabolic jet base \cite{punsly01} also permits gaps that form near the light
cylinder at large angles to the jet axis (``outer gaps''), although numerical studies
(\textit{39})
indicate that the angular distribution of cascade photons developing in a polar gap
is broad enough to explain the emission of gamma rays at large angles to the jet axis.
Another
possibility is that photon-photon pair creation is altogether inefficient, and the seed charges for
the acceleration are pulled out of the accretion flow in which case the positive charges, i.e., protons
and ions, would be accelerated along the jet axis until they lose energy catastrophically by
photo-production of secondaries in interactions with synchrotron photons from the downward
accelerated electrons.

\bibliographystyle{Science}

\begin{thebibliography}{}
\bibitem[16]{notedoppler} In special relativity, the Lorentz and Doppler factors are defined as $\Gamma_{\mathrm{j}}=\left(1-\beta^2\right)^{-1/2}$ 
  and  $\delta=\left[\Gamma_{\mathrm{j}}(1-\beta\cos\theta)\right]^{-1}$, respectively,
  where $\beta$ denotes the dimensionless shock velocity
  and $\theta$ the angle between the line of sight and the direction of the jet, ignoring the cosmological $(1+z)$ factor.  The apparent bolometric luminosity differs from its
  isotropic comoving-frame value by the factor $\delta^4$.
\end{thebibliography}

\clearpage

\begin{figure}
   \centering
      \includegraphics[width=10cm]{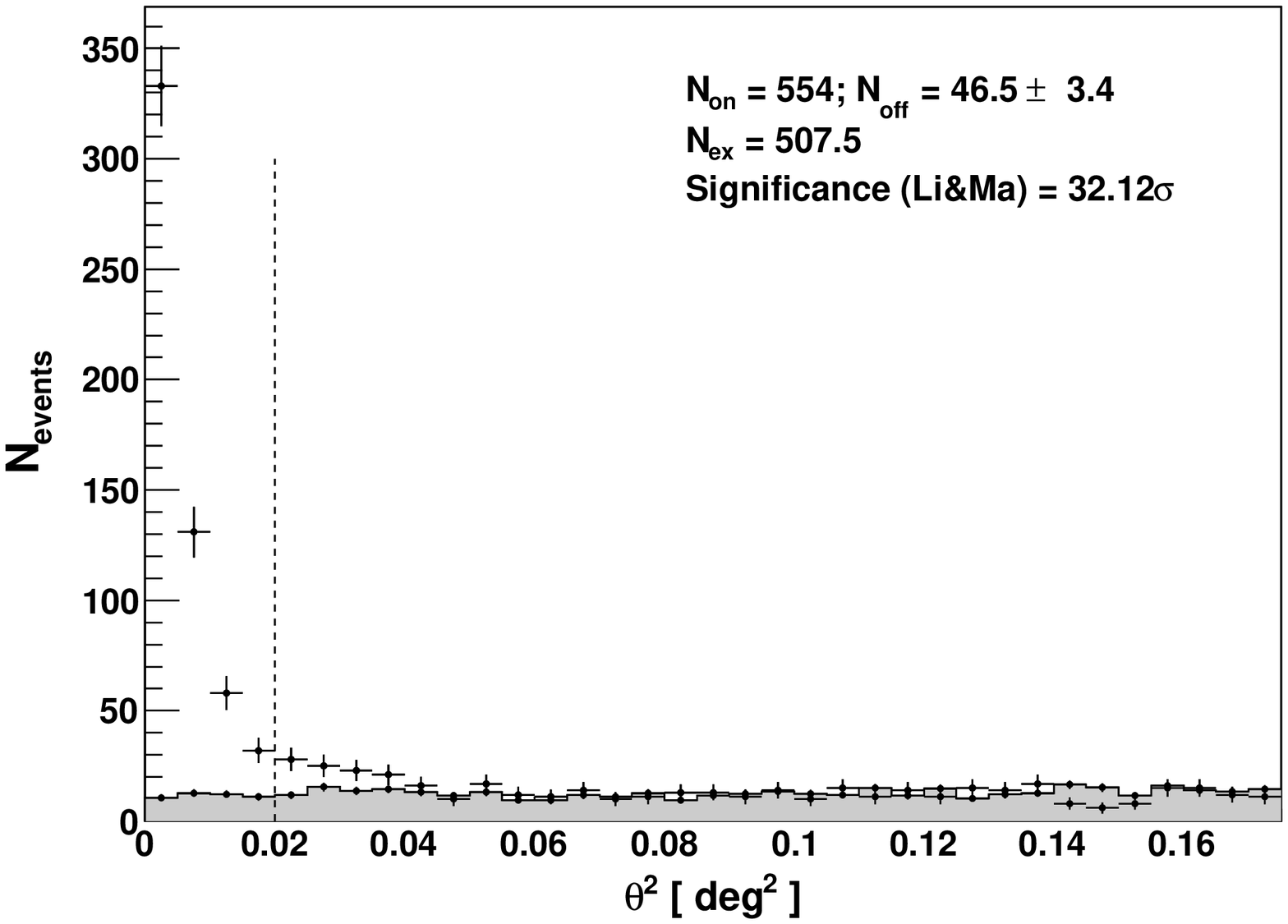}
     \caption{Distribution of the squared angular distance between assumed and reconstructed source 
position ($\theta^2$) of the events in the night of November 12/13$^{\mathrm{th}}$, 2012 above 300\,GeV. 
Black points show the distribution of the 'ON' and the gray shaded area indicates the 'OFF' events. 
This is a the stacked result of individual wobble pointings. 
The number of excess events $N_\mathrm{ex}=N_\mathrm{on}-N_\mathrm{off}$ is calculated in the region from $0$ to the dashed line.}
\label{figS1}
\end{figure}

\begin{figure}
  \centering
  \includegraphics[width=16.0cm]{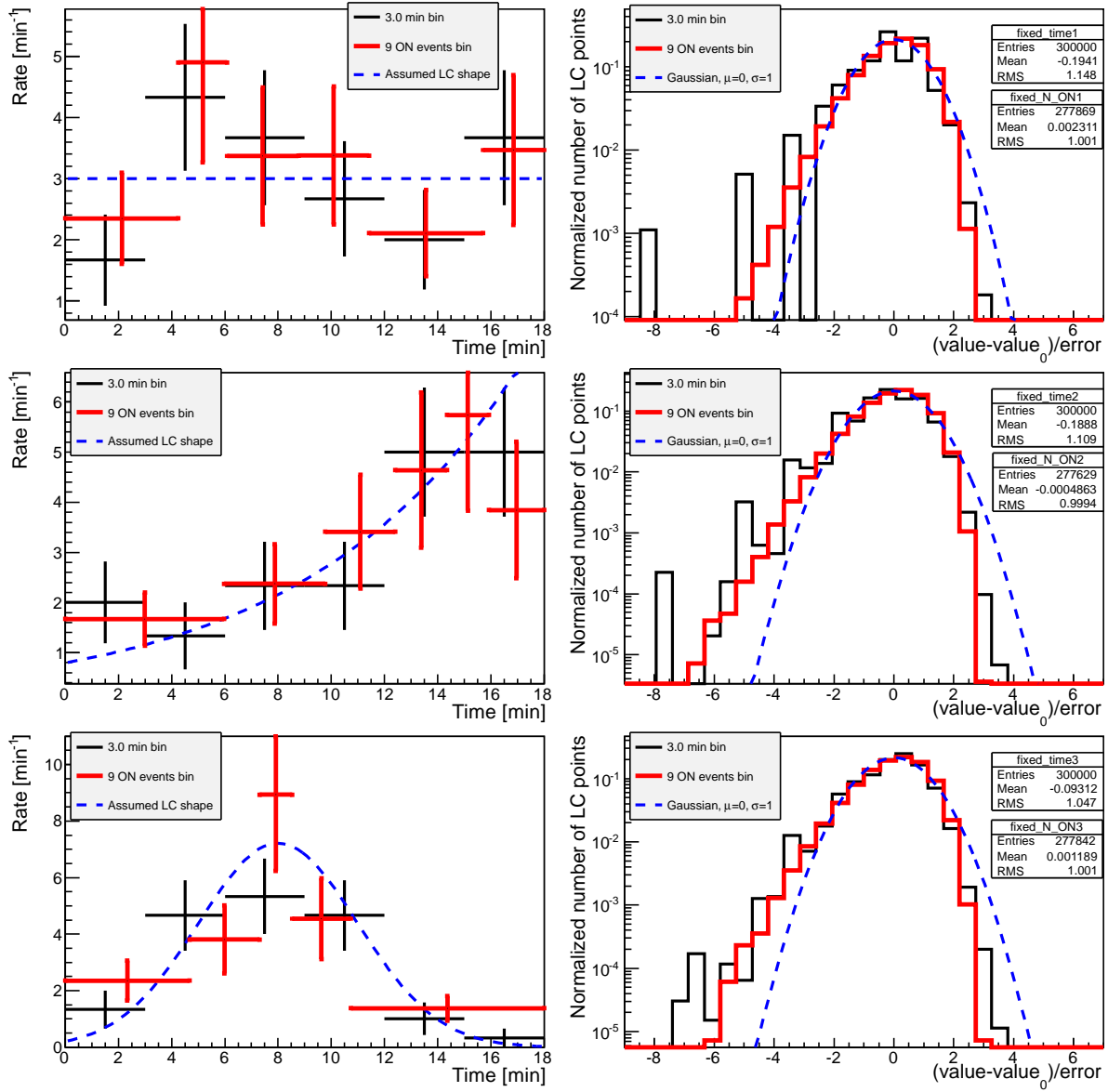}
   \caption{
Comparison of toy Monte Carlo light curves obtained with constant widths of time bins (black) and with fixed 
number of ON events(red) for different assumed shapes of the light curve (dashed blue line): constant 
flux (top panels), 
exponential increase (center panels) and Gaussian peak (bottom panel).
Left panels shows an example light curve for each case, while the right panels show the distributions 
of light curve residua with respect to the assumed shape obtained from 50000 random light curves.
}
\label{figS2}
\end{figure}

\begin{figure}
   \centering
      \includegraphics[width=16cm]{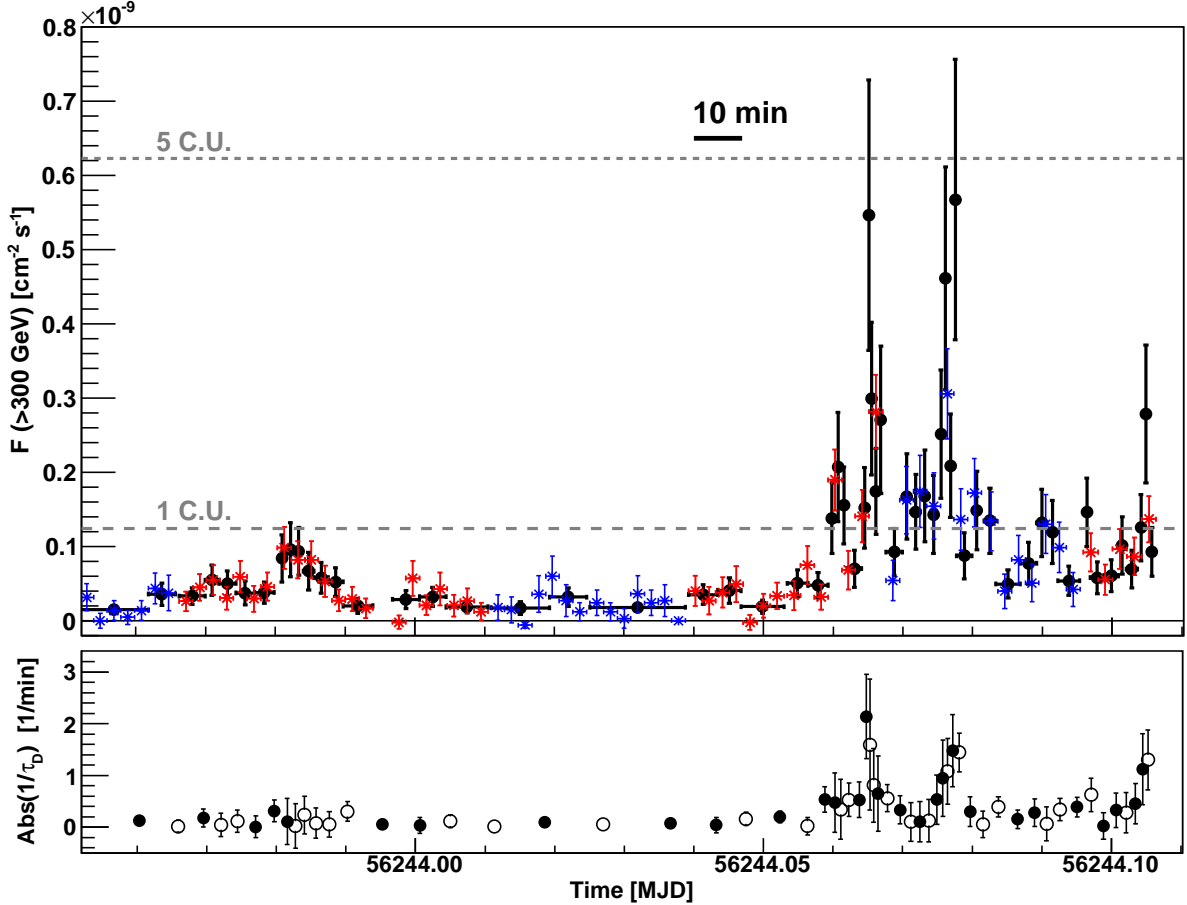}
     \caption{Upper panel: Light curve of IC\,310 as measured in the night of November 13, 2012 above 300\,GeV. 
              The black points show the results obtained from a binning with fixed number of ON events (here: 9 events). The 
              colored light curve presents
              the result from a fixed time binning (here: 3 minutes), red from the data of the wobble pointing 
              with 0.4$^{\circ}$, and blue with 0.94$^{\circ}$ offset, respectively.
              The two gray lines correspond to 1 and 5 times the flux of the Crab Nebula (C.U.) (\textit{25}). 
              Bottom panel: Evolution of the absolute inverse doubling time during the flare computed from the light curve 
              with fixed number of ON events. The  
              doubling time $\tau_\mathrm{D}$ was calculated here by 
              computing the flux difference between each two consecutive points and taking into 
              account the time lapse between the
              two points. Positive doubling times are shown with filled marker and negative doubling times
              with open markers. Vertical error bars show 1 standard deviation statistical uncertainities.
Horizontal error bars in the upper panel show the bin widths.}
\label{figS3}
 \end{figure}

\begin{figure}
   \centering
      \includegraphics[width=7.8cm]{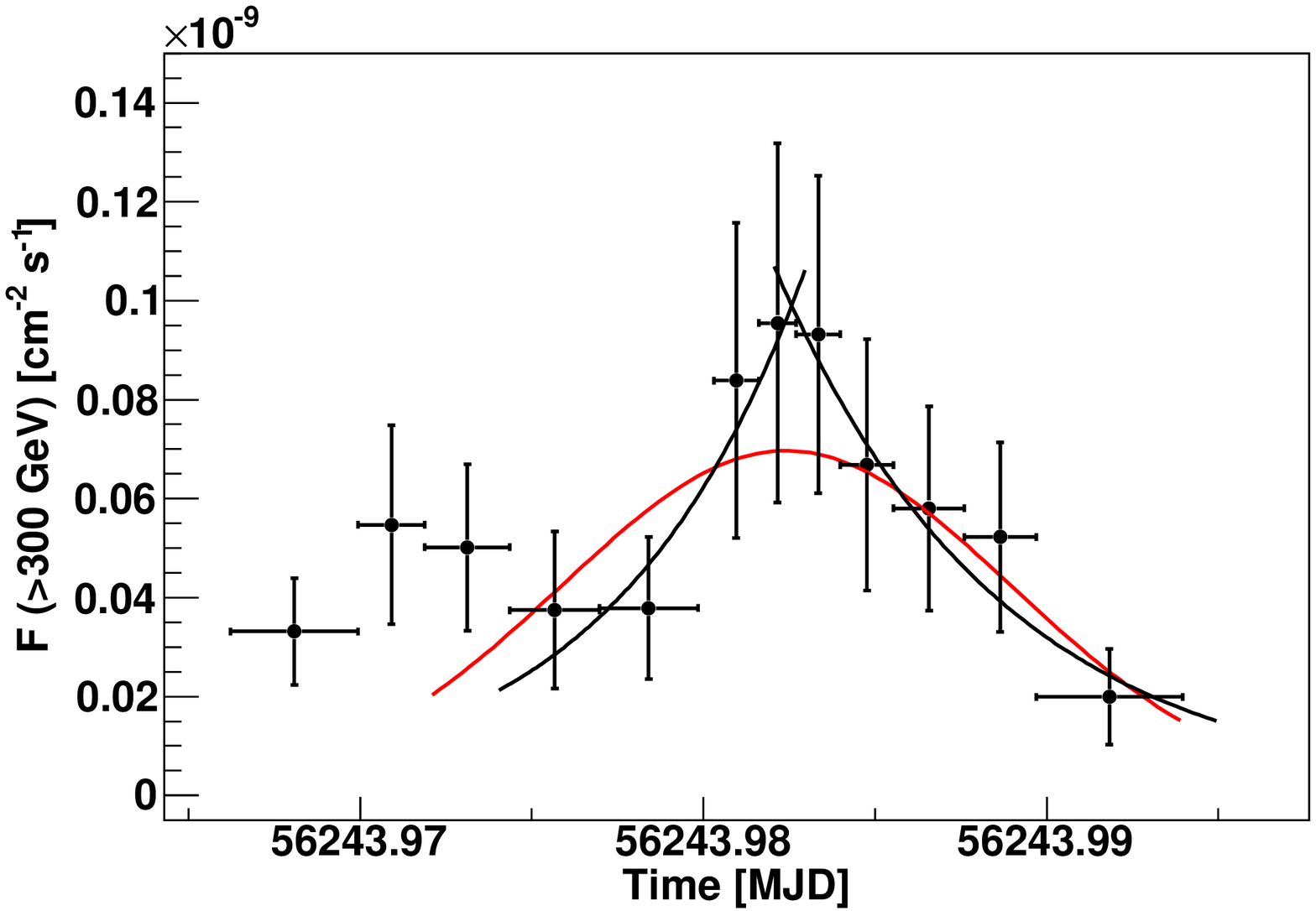}
      \includegraphics[width=7.8cm]{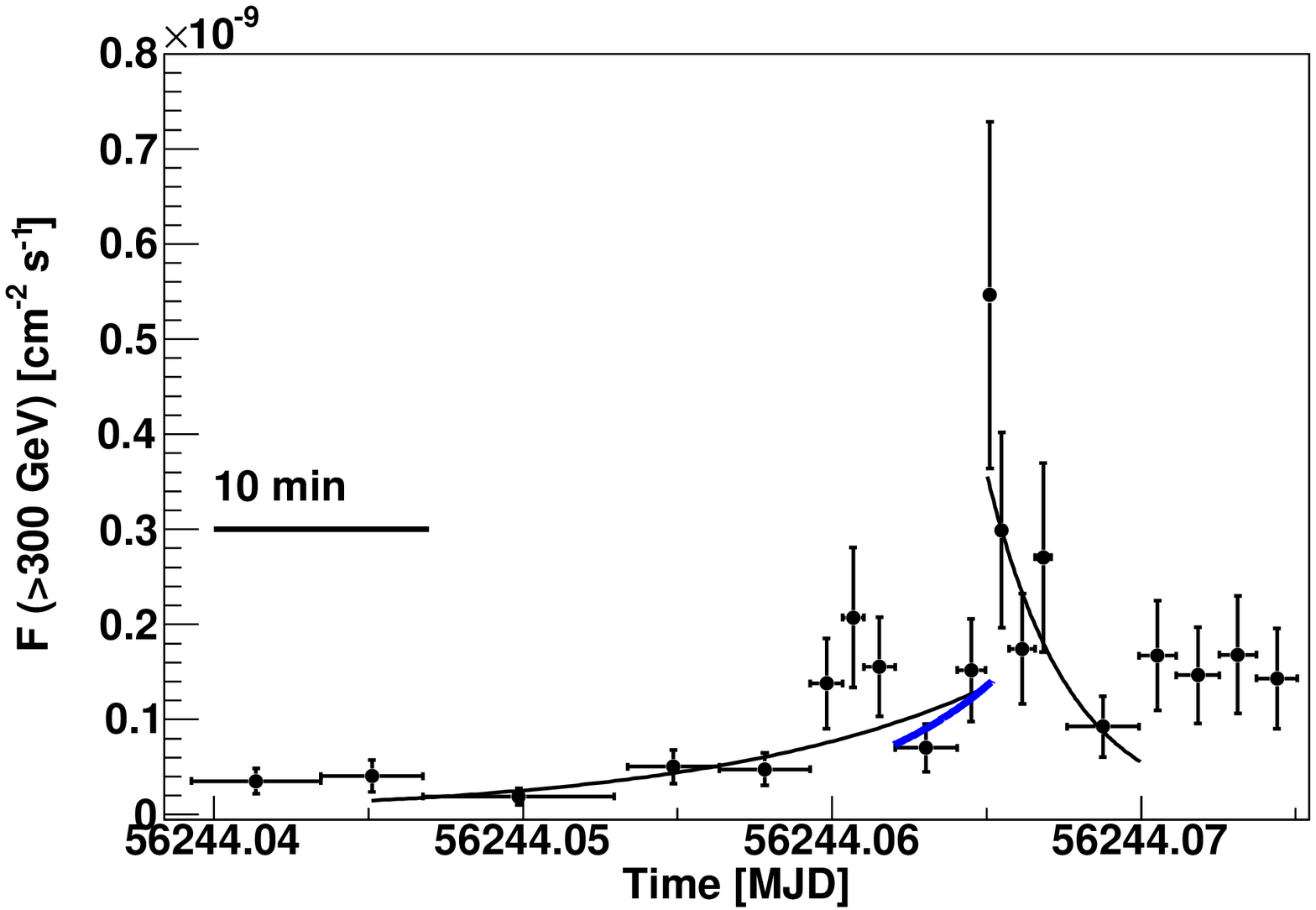}
   \caption{Light curve of IC\,310 of the flare above 300\,GeV in separate time ranges with exponential fits to the rising and decay
edges to the substructures in the light curve (black lines). Left panel: Zoom to the pre flare together with a Gaussian fit (red). Right panel:
Zoom to the first big flare. The blue line shows the slowest doubling time necessary to
explain the raising part of the flare at C.L. of 95\%. Vertical error bars show 1 standard deviation statistical uncertainities.
Horizontal error bars show the bin widths.}

\label{figS4}
    \end{figure}

\begin{figure}
  \centering
  \includegraphics[width=7.5cm]{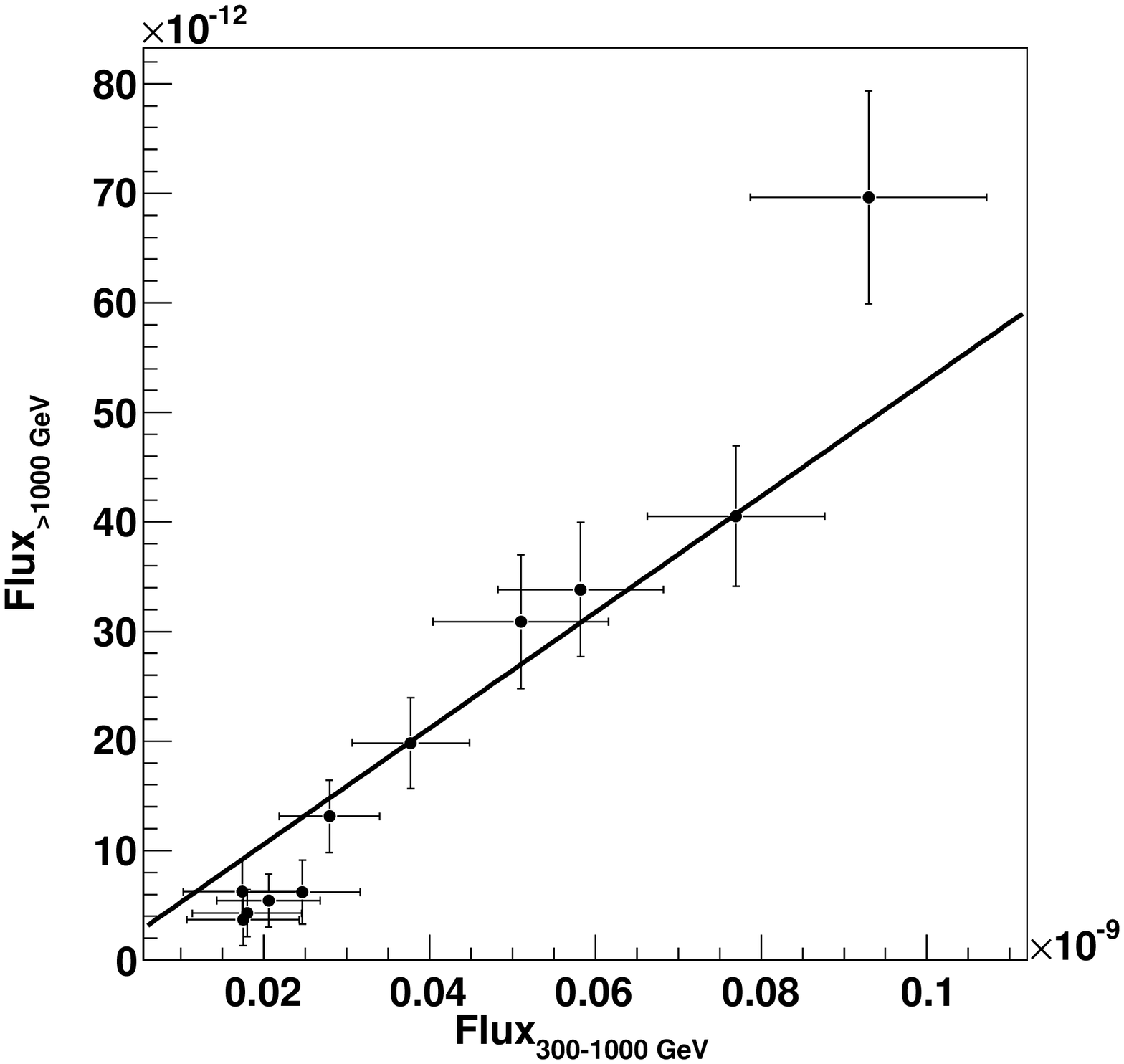}
  \includegraphics[width=7.5cm]{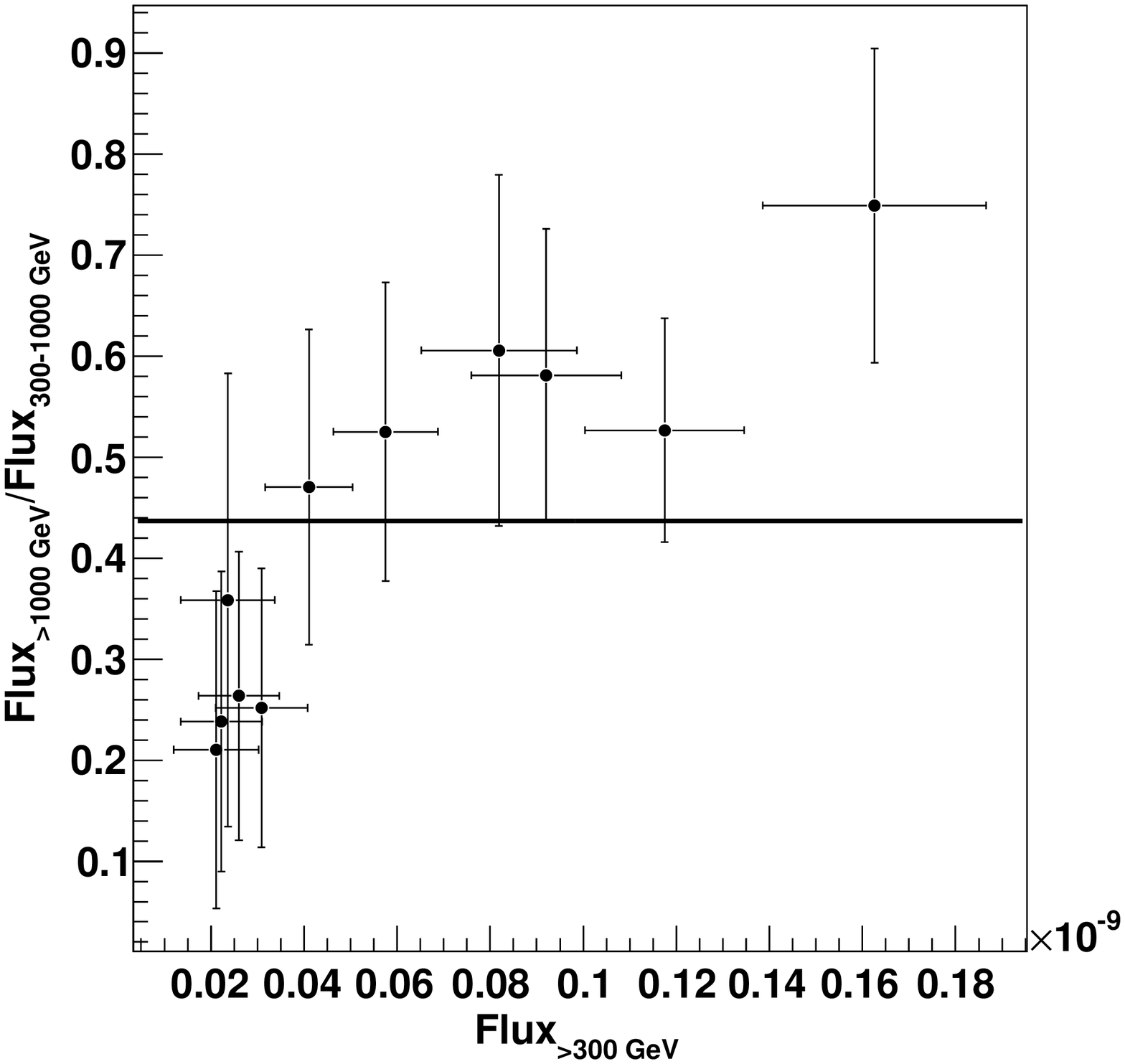}
  \caption{
    Study of spectral variability during the flare using the hardness ratio based on run-wise 
($\sim20$\,min) binned light curves in the low (between 300\,GeV and 1\,TeV) and the high 
energy band (above 1\,TeV).
    Left panel: Flux above 1\,TeV versus flux in the energy range 300\,GeV to 1\,TeV. The linear fit
    has a $\chi^2/$d.o.f. of 10.8/10 corresponding to a probability of 0.37. 
    Right panel: Hardness ratio defined as the ratio between the flux above 1\,TeV and the flux in the energy range 300-1000\,GeV versus the entire flux above 300\,GeV. 
A fit with a constant 
 reveals a $\chi^2/$d.o.f. of 14.3/10 with a probability of 0.16. Error bars show 1 standard deviation statistical uncertainity.}
\label{figS5}
    \end{figure}

\clearpage

\begin{table}
\label{tabS1}     
\centering                        
\begin{tabular}{c c c c }       
\hline
   state  	&  time range 		& $\tau_{\mathrm{D}}=\tau*ln2$ 	& $\chi^2/$d.o.f.\\    
		&  [MJD]      		&  [min] 				&	\\
\hline 
pre flare: 	&           		&       &\\
  rise		& 56243.974--56243.983	& $5.53\pm3.05$	& 0.7982/2		\\	      
  decay		& 56243.982--56243.995	& $6.56\pm1.99$	& 0.822/4		\\	
\hline      
1$^{\mathrm{st}}$ big flare  	&               	&       &     \\
slow rise      & 56244.045--56244.0652	& $9.02\pm2.22$	& 17.15/8 \\
fast rise, 95\% C. L.   & 56244.062--56244.0652	& $<4.88$	& - \\
  decay         & 56244.065--56244.07	& $2.66\pm1.00$	& 3.085/3 \\
\hline                                  
\end{tabular} 
\caption{Fit results from individual substructures in the gamma light curve.} 
\end{table}

\begin{table}
\label{tabS2}     
\centering                         
\begin{tabular}{c c c c c}        
\hline
State    &		&Energy range  &$f_{0}\pm f_{\mathrm{stat}}\pm f_{\mathrm{syst}}$ &$\Gamma\pm\Gamma_{\mathrm{stat}}\pm\Gamma_{\mathrm{syst}}$\\
         &		&[TeV]	       &$\times10^{-12}[\mathrm{TeV}^{-1}\,\mathrm{cm}^{-2}\,\mathrm{s}^{-1}]$&\\
\hline
flare 		   	&obs.  &0.07-8.3  &$17.7\pm0.9\pm2.1$     &$1.90\pm0.04\pm0.15$\\ 
			&intr. &0.07-8.3 &$21.5\pm1.1\pm2.6$     &$1.78\pm0.05\pm0.15$\\  
high 2009/2010        	&obs.  &0.12-8.1 &$4.3\pm0.2\pm0.7$     &$1.96\pm0.10\pm0.20$\\
			&intr. &0.12-8.1 &$5.1\pm0.3\pm0.9$     &$1.85\pm0.11\pm0.20$\\
low 2009/2010     	&obs.  &0.12-8.1 &$0.61\pm0.04\pm0.11$   &$1.95\pm0.12\pm0.20$\\
			&intr. &0.12-8.1 &$0.74\pm0.05\pm0.14$   &$1.81\pm0.13\pm0.20$\\
\hline      
\end{tabular}
\caption{Results of power-law fits of the spectra of the flare as well as previous measurements
 obtained with MAGIC (\textit{28}). Observed as well as intrinsic spectra are given.}
\end{table}

\end{document}